# Tests of Sunspot Number Sequences:
# 1. Using Ionosonde Data

M. Lockwood • C.J. Scott • M.J. Owens • L. Barnard • D.M. Willis

*Abstract*.  More than 70 years ago it was recognised that ionospheric F2-layer critical frequencies [foF2] had a strong relationship to sunspot number.  Using historic datasets from the Slough and Washington ionosondes, we evaluate the best statistical fits of foF2 to sunspot numbers (at each Universal Time [UT] separately) in order to search for drifts and abrupt changes in the fit residuals over Solar Cycles 17–21.  This test is carried out for the original composite of the Wolf/Zürich/International sunspot number [$R$], the new "backbone" group sunspot number [$R_{BB}$] and the proposed "corrected sunspot number" [$R_C$].  Polynomial fits are made both with and without allowance for the white-light facular area, which has been reported as being associated with cycle-to-cycle changes in the sunspot number–foF2 relationship.  Over the interval studied here, $R$, $R_{BB}$, and $R_C$ largely differ in their allowance for the "Waldmeier discontinuity" around 1945 (the correction factor for which for $R$, $R_{BB}$ and $R_C$ is, respectively, zero, effectively over 20 %, and explicitly 11.6 %).  It is shown that for Solar Cycles 18–21, all three sunspot data sequences perform well, but that the fit residuals are lowest and most uniform for $R_{BB}$.  We here use foF2 for those UTs for which $R$, $R_{BB}$, and $R_C$ all give correlations exceeding 0.99 for intervals both before and after the Waldmeier discontinuity.  The error introduced by the Waldmeier discontinuity causes $R$ to underestimate the fitted values based on the foF2 data for 1932–1945 but $R_{BB}$ overestimates them by almost the same factor, implying that the correction for the Waldmeier discontinuity inherent in $R_{BB}$ is too large by a factor of two.  Fit residuals are smallest and most uniform for $R_C$ and the ionospheric data support the optimum discontinuity multiplicative correction factor derived from the independent Royal Greenwich Observatory (RGO) sunspot group data for the same interval.

✉        M. Lockwood
         m.lockwood@reading.ac.uk

1        Department of Meteorology, University of Reading, UK

2        RAL Space, Rutherford Appleton Laboratory, UK

3        also at Centre for Fusion, Space and Astrophysics, University of Warwick, UK

# 1. Introduction

## 1.1 Definitions of Sunspot Numbers

The sunspot number is defined from the well-known formula introduced in its final form (with allowance for observer calibration) by Rudolf Wolf in 1861:

$$R = k \times (10N_G + N_S) \qquad (1)$$

where $N_G$ is the number of sunspot groups, $N_S$ is the number of individual sunspots, and $k$ is the calibration factor that varies with location, instrument and observer (Wolf, 1861). Note that $k$ values for different observers can differ by a factor as large as three (Clette *et al.*, 2015), so accurate estimation of $k$ is absolutely essential to accurate sunspot number derivation. To extend the data series to times before those when both $N_G$ and $N_S$ were recorded systematically, Hoyt and Schatten (1994, 1998) defined the group sunspot number $R_G$ to be

$$R_G = 12.08 \times <k' \times N_G>_n \qquad (2)$$

where $k'$ is the site/observer factor and the averaging is done over the $n$ observers who are available for that day. The factor of 12.08 was designed to make $R$ and $R_G$ values as similar as possible for the more recent data when both $N_G$ and $N_S$ are quantified: specifically it made the mean value of $R_G$ and $R$ the same over 1875–1976. It is well known that $R$ and $R_G$ diverge as one goes back in time. This could be due to real long-term changes in the ratio $N_S/N_G$, but otherwise it would reflect long-term drifts in the calibration of either $R$ or $R_G$ or both.

Note that the observer calibration factors $k$ in Equation (1) are relative and not absolute, independently determined factors, being defined for an interval $T$ as $<R_W/R_O>_T$ where $R_W$ is Wolf's sunspot number from a central reference observatory (for which $k$ is assumed to be constant and unity) and $R_O$ is that derived by the observer in question. Because the $k$-values in the modern era vary by a factor of up to three with location, equipment, and observer, all of which change over time, in general we must expect $k$-values for historic observations to have the potential to vary with time by at least this factor, and probably more (Shapley, 1947) The same is true for the $k'$-factors used in the compilation of $R_G$.

Another point about the definitions of $R$ and $R_G$ is that they both inevitably require subjective decisions to be made by the observer to define both spots and groups of spots on the visible solar disk. Hence observer bias is a factor. Furthermore, the nature of the subjective decisions required has changed with observing techniques and as new guidelines and algorithms were established to try to homogenise the observations and may even have changed for one observer over their lifetime. Some of the effects of these subjective decisions are subsumed into the *k*-values but others are not because they change with time. Also the assumption that $k = 1$ at all times for the reference station must be challenged. With modern digital white-light images of the solar disk, it is possible to deploy fixed and objective algorithms to deconvolve all instrumental effects and define what constitutes a spot and what constitutes a group of spots. For such data, the main subjective decision needed is as to when obscuration by clouds, mists, or atmospheric aerosols is too great for a given site: with sufficient observatories around the globe, unobscured observations are always available from some locations but a decision is needed as to which to employ to ensure that average sunspot numbers are not influenced by inclusion of data from observatories suffering from partial obscuration. Prior to the availability of digital images, photographic plates are available. For these, there are additional considerations about image contrast, telescope focus, scattered-light levels, image exposure time, and resolution (collectively giving net observer acuity).

The most important subjective decision required of observers is what constitutes a group, which is of crucial importance given the weighting given to $N_G$ in Equation (1). However, there are other subjective decisions that influence both $N_S$ in $N_G$. For example, sunspots must be distinguished from pores, which are smaller than sunspots (typically 1–6 Mm, compared to 6–40 Mm for sunspots) and sometimes, but not always, develop into sunspots (Sobotka, 2003). Their intensity range overlaps with that for sunspots, at their centre being $0.2 I_{ph} - 0.7 I_{ph}$ (where $I_{ph}$ is the mean photospheric intensity) compared to the $0.05 I_{ph} - 0.3 I_{ph}$ for sunspots. In sufficiently high-resolution images, sunspots and pores are distinguished by the absence of a sunspot penumbra around pores (although some pores show unstable filamentary structures that can be confused with a sunspot penumbra).

The original photographic glass plates acquired by the Royal Observatory, Greenwich and the Royal Greenwich Observatory (collectively here referred to as "RGO") during the interval 1918 – 1976 still survive. These are currently stored in the "Book Storage Facility" in South Marston, near Swindon, UK, as part of the Bodleian Libraries, Oxford. The RGO glass plates for the earlier interval 1873 – 1917 are thought to have been destroyed during the First World War. However, contact prints (photographs) were made of some, but certainly not all, of these earlier glass plates before they were lost (in particular, plates not showing any obvious sunspots were not copied). The fraction of days for which there are no contact prints is considerably higher before 1885 (Willis, Wild, and Warburton, 2015). The extant contact prints form part of the official RGO Archives, which are stored in the Cambridge University Library (Willis *et al.*, 2013a; 2013b)

Most of the information available before 1874 is in the form of sketches of the solar disk and/or tabulated sunspot and/or sunspot group counts compiled by observers using a telescope. (However, note that even after 1918 sunspot numbers were frequently compiled without the use of photographic images). It is for these non-photographic records that the subjective nature of sunspot number data is greatest and the $k$- and $k'$-factors are most uncertain and least stable. Because observers will have used different criteria to define both spots and spot groups (and even a given observer's criteria may have changed with time) and because observer acuity varies from observer to observer and with time, intercalibrations of data are required (*e.g.,* Chernosky and Hagan, 1958). All long-term sunspot-number data sequences are therefore an observational composite: this is true of the much-used original Wolf/Zurich/International sunspot-number data sequence (version 1 of the International Sunspot Number, here termed $R$) as published by Solar Influences Data Analysis Center (SIDC, the solar physics research department of the Royal Observatory of Belgium) and hence of all sunspot series based on $R$ with corrections for known or putative discontinuities, for example, the corrected sequence [$R_C$] suggested by Lockwood *et al.* (2014). This is equally the case for the new (second) version of the Wolf/Zurich/International composite recently published by SIDC, the sunspot-group number [$R_G$] (Hoyt and Schatten, 1994, 1998), and the "backbone" group number data series [$R_{BB}$] proposed by Svalgaard and Schatten (2015).

To compile the backbone series [$R_{BB}$] a primary observation source is selected to cover a given interval and the quality of other observers is judged by how well they correlate with the chosen backbone. Sequences put together this way were then "daisy-chained" using intercalibrations of the segments from the interval of overlap between the two to give $R_{BB}$. Obviously, the choices of which data sequences are chosen to be backbones are critical. It is important to note that the intercalibration of observers should be done on a daily basis because sunspot groups can appear and disappear in as little as one day (Willis, Wild, and Warburton, 2015). Cloud cover means that observers do not, in general, make observations on the same days, and this will introduce errors if intercalibration is carried out on annual, or even monthly, means of the two incomplete data sets. Hence intercalibrations carried out on daily data, such as those by Usoskin *et al.* (2016) are much more reliable than those done on annual means, as used to generate $R_{BB}$.

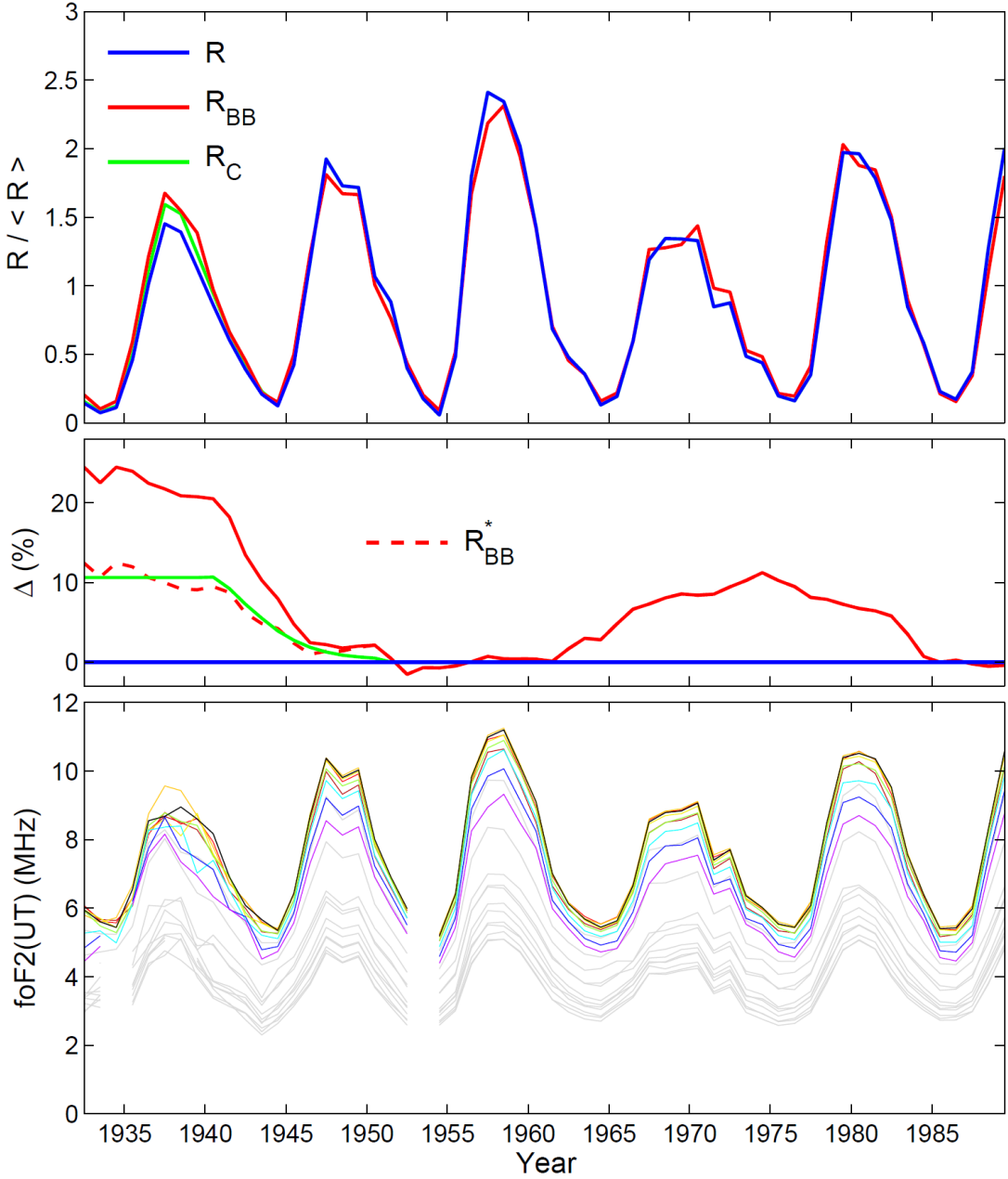

**Figure 1.** (a) Annual mean sunspot data time series used in this article: the old (version 1) SIDC composite of Wolf/Zürich/International sunspot number [$R$] (in blue), the new Backbone sunspot group number [$R_{BB}$] (in red); the corrected sunspot number [$R_C$] (for a best-fit factor of 11.6%, in green). (b) Eleven-year running means of percentage deviations of normalised variations from $R$: (in red) for $R_{BB}$, $\Delta_{BB} = 100 \times \{(R_{BB}/\langle R_{BB}\rangle) / (R/\langle R\rangle) - 1\}$, (in green) for $R_C$, $\Delta_C = 100 \times \{(R_C/\langle R_C\rangle) / (R/\langle R\rangle) - 1\}$. The red dashed line is for $R_{BB}^*$, which is $R_{BB}$ with application of the optimum correction for the Waldmeier discontinuity found in this article. $R_{BB}^*$ is 12% smaller than $R_{BB}$ for all times before 1945. (c) Annual mean F2 layer critical frequencies [foF2], measured on the hour for each of 24 Universal Times (UT). Coloured lines are for the nine UTs at which the sequences give a correlation coefficient exceeding 0.99 when fitted to all of $R$, $R_{BB}$, and $R_C$ (the colours are given in Figures 3 and 4: note that the values for local Noon are in black). Grey lines are for the other 15 UTs that do not meet this criterion.

Figure 1a shows the sequences of $R$, $R_C$, and $R_{BB}$ for the interval analysed in the present article (in blue, green, and red, respectively). As discussed in Article 2 (Lockwood *et al.*, 2015a), while the differences over the interval in Figure 1 are relatively minor, they continue to grow as one goes back in time.

The differences in sunspot numbers caused by the subjective decisions required of the observers, by their instrumentation performance, and by local cloud and atmospheric air-quality conditions, make definitive calibration of individual observers extremely difficult, if not impossible. The term "daisy chaining" refers to all methods for which the calibration is passed from one data segment to the next using a relationship between the two, derived from the period of overlap. Usually this relationship has been obtained using some form of regression fit. However, as noted by Lockwood *et al.* (2006) and by Article 3 in this series (Lockwood *et al.*, 2015b), there is no definitively-correct way of making a regression fit and tests of fit residuals are essential to ensure that the assumptions made by the regression have not been violated as this can render the fit inaccurate and misleading for the purposes of scientific deduction of prediction. Article 3 shows that large intercalibration errors from regression techniques (>30 %) can arise even for correlations exceeding 0.98 and that no one regression method is always reliable in this context: use of regression frequently gives misleading results that amplify the amplitude of Solar Cycles in data from lower-acuity observers. The problem with daisy-chaining is that any errors (random and systematic) in the relationship will apply to all data before that error (assuming modern data are the most accurate) and if there is a systematic bias, the systematic errors are in the same sense and will compound, such that very large deviation can result by the start of the data sequence. The intercalibrations also depend upon subjective decisions about which data to rely on most, over which intervals to intercalibrate, and on the sophistication and rigour of the chosen statistical techniques. Thus daisy-chaining of different data is a likely source of spurious long-term drift in the resulting composite. Most observational composites until now have been assembled using some form of daisy chaining and so are prone to the propagation of errors (this is certainly true of $R$, $R_{BB}$, and $R_C$). An important exception, which avoids both daisy chaining and regression, is the new composite of group numbers $R_{UEA}$ assembled by Usoskin *et al.* (2015) who compared data probability distribution functions for any interval with those for a fixed standard reference set (the RGO data after 1920 were used). In particular, they used the fraction of observed days that revealed no spots to obtain a calibration rather than passing the calibration from one data segment to the next. These authors assumed that the calibration of each observer remained constant over their observing lifetime; however, their method could be refined and applied to shorter intervals to allow for the drift in each observer's calibration factor over time.

For a number of reasons, it is highly desirable that sunspot data series are compiled using only sunspot observations. Other data, such as geomagnetic observations, the frequency of occurrence of low-latitude aurorae, or cosmogenic isotope abundance measurements, correlate on a range of timescales but it cannot be assumed that the regression coefficients are independent of timescale. Hence using such data to calibrate the sunspot data on centennial timescales may introduce long-term differences. An example, in the context of the present article, is that ionospheric F-layer critical frequencies [foF2], and sunspot numbers correlate very well on decadal timescales. However, it has been proposed that anthropogenic warming of the troposphere by greenhouse gases and the associated cooling of the stratosphere, mesosphere, and thermosphere cause lowering of ionospheric layers (through atmospheric contraction) and could potentially influence ionospheric plasma densities and critical frequencies (Roble and Dickinson, 1989; Rishbeth, 1990; Ulich and Turunen, 1997). Furthermore, any such effects will be complicated by changes in the local geomagnetic field (Cnossen and Richmond, 2008). If sunspot calibration were to be based on foF2 values such effects, if present, would not be apparent because it would be included in the sunspot-number intercalibrations, and the sunspot data sequence would contain a spurious long-term drift introduced by the atmospheric and geomagnetic effects. This is just one of many potential examples where using ionospheric data to calibrate sunspot data could seriously harm ionospheric studies by undermining the independence of the two datasets. However, note that these studies are also damaged if an incorrect sunspot data series is used.

It must always be remembered that sunspot numbers have applications only because they are an approximate proxy indicator of the total magnetic flux threading the photosphere and hence can be used to estimate and reconstruct terrestrial influences such as the received shortwave Total Solar Irradiance (TSI) and UV irradiance (Krivova, Balmaceda, and Solanki, 2007; Krivova *et al.*, 2009, respectively), the open solar magnetic flux (Solanki, Schüssler, and Fligge, 2000; Lockwood and Owens, 2014a), and hence also the near-Earth solar-wind speed (Lockwood and Owens, 2014b), mass flux (Webb and Howard, 1994), and interplanetary magnetic-field strength (Lockwood *et al.*, 2014a). Sunspot numbers also provide an indication of the occurrence frequency of transient events, in particular coronal mass ejections (Webb and Howard, 1994; Owens and Lockwood, 2012) and the phase of the decadal-scale sunspot cycle is used to quantify the tilt of the heliospheric current sheet (Altschuler and Newkirk, 1969; Owens and Lockwood, 2012) and hence the occurrence of fast solar-wind streams and co-rotating interaction regions (Smith and Wolf, 1976; Gazis, 1996). Because all of the above factors influence the terrestrial space environment, sunspot numbers are useful in providing an approximate quantification of terrestrial space-weather and space-climate phenomena and hence it is vital that the *k*-factor intercalibrations inherent in all sunspot number composites mean that their centennial drifts correctly reflect trends in the terrestrial responses.

From the above arguments, we do not advocate using correlated data to calibrate sunspot numbers, but we do think it important to evaluate any one sunspot number data sequence against the trends in terrestrial effects because it is these effects that give sunspot numbers much of their usefulness.

**1.2 The "Waldmeier Discontinuity"**

In this article, we look at the long-term relationship between the sunspot number data sequences *R*, $R_C$, and $R_{BB}$ and the ionospheric F2-region critical frequency [foF2] for which regular measurements are available since 1932. This interval is of interest as there has been discussion about a putative inhomogeneity in the calibration of sunspots data series around 1945 that has been termed the "Waldmeier discontinuity" (Svalgaard, 2011; Aparicio *et al.*, 2012, Cliver *et al.*, 2013). This is thought to have been caused by the introduction of a weighting scheme for sunspot counts according to their size and a change in the procedure used to define a group (including the so-called "evolutionary" classification that considers how groups evolve from one day to the next); both changes that may have been introduced by the then director of the Zürich observatory, Max Waldmeier, when he took over responsibility for the production of the Wolf sunspot number in 1945. Note that these changes affect both sunspot numbers and sunspot-group numbers, but not necessarily by the same amount. Svalgaard (2011) argues that these corrections were not applied before this date, despite Waldmeier's claims to the contrary. By comparison with other long time series of solar and solar-terrestrial indices, Svalgaard makes a compelling case that this discontinuity is indeed present in the data. Svalgaard argues that sunspot number values before 1945 need to be increased by a correction factor of 20 %, but it is not clear how this value was arrived at beyond visually inspecting a plot of the time variation of the ratio $R_G/R$ (neglecting low *R* values below an arbitrarily chosen threshold as these can generate very large values of this ratio). Note that this assumes that the correction required is purely multiplicative, *i.e.* that before the discontinuity the corrected value $R' = f_R \times R$ (and Svalgaard estimates $f_R = 1.2$) to make the pre-discontinuity values consistent with modern ones.

Lockwood, Owens, and Barnard (2014) studied fit residuals when *R* is fitted to a number of corresponding sequences. These were: i) the independent sunspot-group number from the Royal Greenwich Observatory (RGO) dataset; ii) the total group area data from the RGO dataset; and iii) functions of geomagnetic activity indices that had been derived to be proportional to sunspot numbers. For each case, they studied the difference between the mean residuals before and after the putative Waldmeier discontinuity and quantified the probability of any one correction factor with statistical tests. These authors found that the best multiplicative correction factor [$f_R$] required by the geomagnetic data was consistent with that for the RGO sunspot-group data but that the correction factor was very poorly constrained by the geomagnetic data. Because both the sample sizes and the variances are not the same for the two data subsets (before and after the putative discontinuity), these authors used Welch's t-test to evaluate the probability p-values of the difference between the mean fit residuals for before and after the putative discontinuity. This two-sample t-test is a parametric test that compares two independent data samples (Welch, 1947). It was not assumed that the two data samples are from populations with equal variances, so the test statistic under the null hypothesis has an approximate Student's t-distribution with a number of degrees of freedom given by Satterthwaite's approximation (Satterthwaite, 1946). The distributions of residuals were shown to be close to Gaussian and so application of nonparametric tests (specifically, the Mann–Whitney U (Wilcoxon) test of the medians and the Kolmogorov–Smirnov test of the overall distributions) gave very similar results. These tests yielded a correction factor of 11.6 % ($f_R$ = 1.116) with an uncertainty range 8.1–14.8 % at the 2σ level. The probability of the factor being as large as the 20 % estimated by Svalgaard (2011) was found to be miniscule (1.6 ×10$^{-5}$). Lockwood, Owens, and Barnard (2014) carried out these tests in two ways. The "before" period was 1874–1945 (*i.e.* all of the prior RGO data were used) in both cases but two "after" periods were used: 1945–2012 and 1945–1976. The former uses data from both RGO and Solar Optical Observing Network (SOON), with some data gaps that are filled using the "Solnechniye Danniye" (Solar Data, SD) Bulletins issued by the Pulkovo Astronomical Observatory in Russia. These data need to be intercalibrated with the RGO data (for example the RGO and SD records were photographic whereas the SOON data are based on sketches) (Foukal, 2013). In the second analysis, for the shorter "after" interval, only the RGO data were used.

In relation to this analysis by Lockwood, Owens, and Barnard (2014), it has been argued that the RGO data are not homogeneous, particularly before about 1915 (Clette *et al.*, 2015; Cliver and Ling, 2015). To be strictly rigorous, the RGO count of the number of sunspot groups on the solar disk is inhomogeneous essentially by definition, since this count is based on information derived from photographs acquired at different solar observatories, which use different solar telescopes, experience different seeing conditions, and employ different photographic processes (Willis *et al.*, 2013a; 2013b). With this rigorous definition, the RGO count of the number of sunspot groups is also inhomogeneous after 1915. It can be shown that the RGO count of the number of sunspot groups in the interval 1874 – 1885 behaves as a "quasi-homogeneous" time series (Willis, Wild, and Warburton, 2015) but the correct decisions have to be taken about how to deal with days of missing data. Moreover, changes in the metadata do not appear to invalidate the integrity of the time series. The stability of the RGO sequence calibration is of relevance here because any drift in the RGO group data could, it has been argued, be at least part of the reason why Lockwood, Owens, and Barnard (2014) derived a lower correction factor for the Waldmeier discontinuity than Svalgaard (2011). The argument is that, because they used all of the RGO data, extending back to 1874, this may have introduced some poorly calibrated data. In the present article, as well as studying the relationship to ionospheric data, we repeat the analysis of Lockwood, Owens, and Barnard (2014) but using shorter intervals and RGO data only; namely, 1932–1945 for the "before" interval and 1947–1976 for the "after" interval. The choice of 1932 is set by the availability of ionospheric data that can be used to make the corresponding tests (the results of which are therefore directly comparable with the tests against the RGO data presented here) but also 1932 is well after the interval of any postulated RGO data calibration drift. The shorter periods mean fewer data points, which necessarily broadens the uncertainty band around the optimum correction-factor estimates.

Figure 1b shows the fractional deviations of the $R_{BB}$ and $R_C$ variations from the commonly-used old version of the international sunspot number [$R$] Because $R_{BB}$ is a group number, whereas $R$ and $R_c$ are Wolf sunspot numbers, we compare them by normalising to their averages over the interval 1932–1976. The percent deviation of normalised $R_{BB}$ is then

$$\Delta_{BB} = 100\times\{(R_{BB}/<R_{BB}>) - (R/<R>)\} / (R/<R>) = 100\times\{(R_{BB}/<R_{BB}>) / (R/<R>) - 1\} \quad (3)$$

$\Delta_{BB}$ is shown by the red line in Figure 1(b). The equation corresponding to equation (3) for $R_C$ is used to compute $\Delta_C$, which is shown by the green line. To illustrate the long-term trends in the calibration, 11-year running means of both $\Delta_{BB}$ and $\Delta_C$ are presented. Because of the 11-year smoothing, the correction applied for 1945 and before in the case of $R_C$ appears as a ramp over the interval 1939–1950. To compile $R_{BB}$, various "backbone" sequences (assumed to be of constant and known $k$) were used with intercalibrations devised by the authors, rather than applying a fixed correction to $R$. It can be seen that the net result is that (over the 11-year interval over which the Waldmeier discontinuity has an effect in these smoothed data) $R_{BB}$ changes by over 20%, relative to $R$. Therefore the correction for the Waldmeier discontinuity inherent in $R_{BB}$ is slightly larger that proposed explicitly by Svalgaard (2011), which Lockwood, Owens, and Barnard (2014) found to be too large by a factor of almost two and to have a probability p-value of $< 10^{-4}$.

The red dashed line in Figure 1 shows the corresponding deviation for $R_{BB}*$, which is $R_{BB}$ with application to 1945 of a 12 % correction to allow for an overestimation of the Waldmeier discontinuity in the compilation of $R_{BB}$. It can be seen that this correction, which will be derived in the present article, brings $R_{BB}$ broadly in line with $R_C$ for the interval studied here. (Note that $R_C$, by definition, is the same as $R$ after the Waldmeier discontinuity but $R_{BB}$ differs from them because it contains some corrections to the Locarno data, which was used as the standard reference ($k = 1$) for much of this interval – those corrections will also be tested in the present article).

Lastly, we note that the corrections proposed by both Svalgaard (2011) and Lockwood, Owens, and Barnard (2014) assume that the corrected values are proportional to the uncorrected ones so that a single multiplicative factor can be used (*i.e.* $R' = f_R \times R$ ). However, Article 3 in this series (Lockwood *et al.*, 2015b) shows that this assumption can be very misleading and Article 4 (Lockwood, M., Owens, M.J., and Barnard, 2015) carries out a number of tests assuming linearity but not proportionality by also allowing for a zero-level offset, δ (*i.e.* $R' = f_R \times R + δ$).

1.**3  Ionospheric F-Region Critical Frequency**

Because foF2 is the largest ordinary-wave mode HF radio frequency that can be reflected by the ionosphere at vertical incidence, it is where the pulse time-of-flight (and hence virtual reflection height) goes to infinity and hence is readily scaled from ionograms generated by ionosondes (vertical sounders with co-located transmitter and receiver). Under the "spread-F" condition, which at middle latitudes occurs predominantly at night, echoes at frequencies above foF2 can be received, caused by reflections off ionospheric plasma irregularities; however, rules for scaling foF2 under these conditions were soon established under international standards (*e.g.* Piggott and Raver, 1961) and foF2 can be readily scaled from the asymptotic limit of the lower edge of the spread in the ionogram trace. Other problems, such as external radio interference, can make the trace hard to define at all frequencies. These problems are greater if transmitter power is low (although much lower powers can be used if advanced pulse-coding techniques are deployed). The main instrumental uncertainty is the accuracy of the transmitter carrier-wave frequency at the relevant point of each frequency sweep, and this varies with the manufacture of the ionosonde in use. Most of the time, especially at middle latitudes during the day, foF2 is a straightforward, objective measurement.

Regular monitoring of foF2 values began in the early 1930s such that by the mid-1940s a whole solar cycle had been observed at several sites, notably Slough in England and Washington, DC in the USA, allowing evaluation of the foF2−$R$ relationship (*e.g.* Allen, 1948). Several authors noted the hysteresis effect whereby the relationship can be slightly different during the rising phase of the cycle than during the falling phase (*e.g.* Ostrow and PoKempner, 1952; Trísková and Chum, 1996; Özgüç, Ataç, and Pektaş, 2008). Furthermore it was noted that, in general, the foF2−$R$ relationship varied from solar cycle to solar cycle (Ostrow and PoKempner, 1952; Smith, and King, 1981; Ikubanni *et al.,* 2013). An example showing some hysteresis and cycle-to-cycle change in data from Washington, DC, USA is presented in Figure 2. In this plot, noon data for Cycle 17 are scaled from the temporal variation of monthly means given by Phillips (1947), and for Cycles 18 and 19 the data are the monthly medians downloaded from the Space Weather Services (SWS, formerly known as IPS) database in Australia (URL given near the end of section 2). These datasets cover 1933.5–1947.5 and 1939–1968, respectively, giving an overlap period of 1939–1947.5, over which interval the two agree so closely that they are almost identical (correlation coefficient r > 0.999) indicating that the two datasets have a common provenance. Both the foF2 and $R$ data shown in Figure 2 are 12-point running averages of monthly data. Figure 2 reproduces the evolution in $R$−foF2 space for two Solar Cycles, as presented by Ostrow and PoKempner (1952) and extends it to a third solar cycle. The lines show best-fit third-order polynomials for the three cycles. From the fitted lines over the range 5.0< foF2≤10.5MHz, the average of the ratio of the values of $R$ for a given foF2 for Cycle 17 to Cycle 18 is 1.316. For Cycles 18 and 19 this ratio is 1.075. If the foF2−$R$ relationship were to be actually the same for these three cycles, this would yield that $R$ in Cycle 17 was 31.5 % low compared to Cycle 18 and that Cycle 18 was, in turn, 7.5 % low compared to Cycle 19. Thus this would imply a 41.5 % drift in the calibration of $R$ in just three Solar Cycles. Clette *et al.* (2015) used the $R$−foF2 plots of the Washington data for Cycles 17 and 18, as published by Ostrow and PoKempner (1952), to attribute all of the change between them to the Waldmeier discontinuity in $R$ and, indeed, this will have made some contribution.

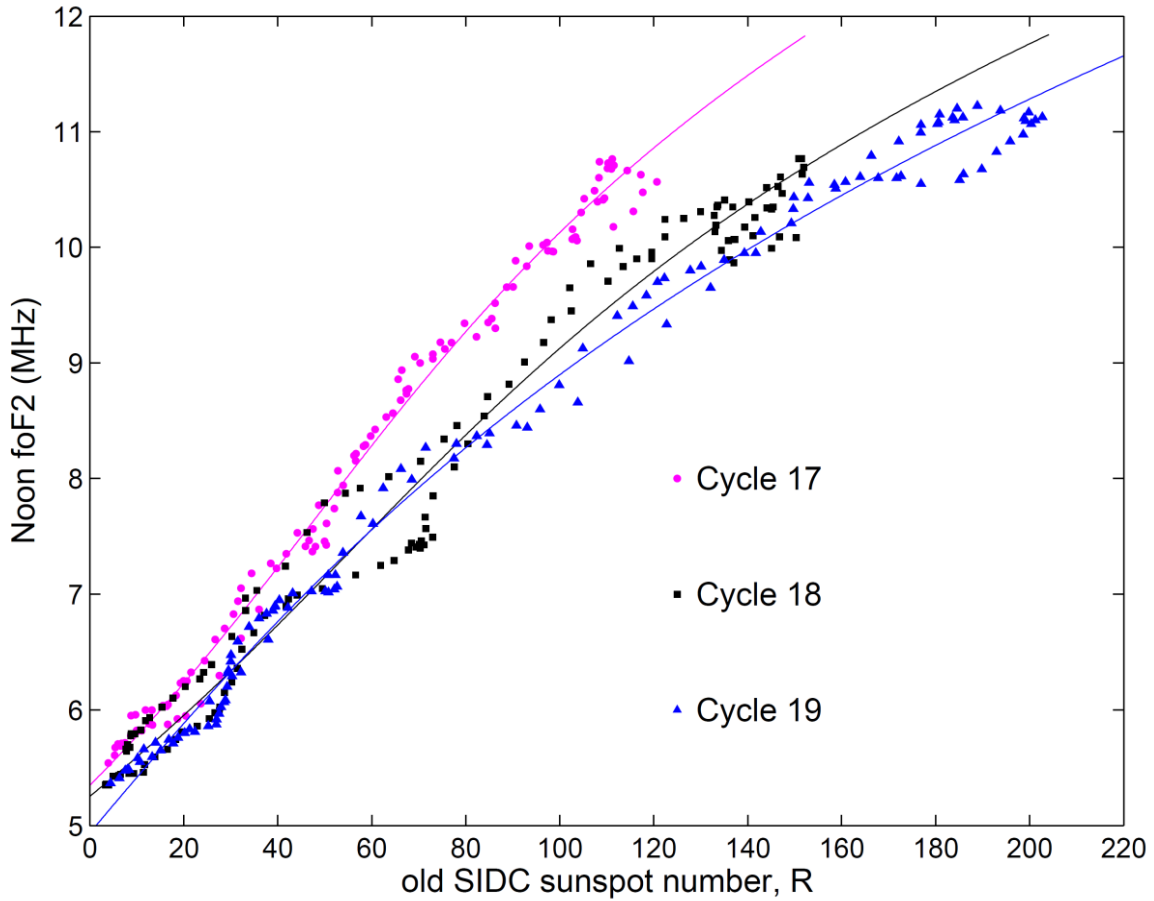

**Figure 2.** Scatter plots of noon foF2 values measured at Washington, DC, USA as a function of sunspot number [$R$], for: (mauve dots) 1933.5–1944 (Cycle17); (black squares) 1944–1954.5 (Cycle 18); and (blue triangles) 1954.5–1964.5 (Cycle 19). Data are 12-point running means of monthly data. The lines are third-order-polynomial fits in each case.

However, there are two issues that show these data cannot, on their own, be used to quantify the correction factor required for the Waldmeier discontinuity (or give evidence to support an independent estimate):

i). Comparison with data from Washington for Cycle 19 shows that the drift in the foF2–$R$ relationship continued after the Waldmeier discontinuity (giving the 7.5% difference between Cycles 18 and 19 in Figure 2) ii). Smith and King (1981) studied the changes in the foF2–$R$ relationship at a number of stations (at times after the Waldmeier discontinuity). For all of the stations that they studied, these authors found that foF2 varied with the total area of white-light faculae on the Sun, as monitored until 1976 by the Royal Greenwich Observatory, as well as with sunspot number. Furthermore, these authors showed that the sensitivity to the facular effect was a strong function of location and that, of the six stations that they studied, it was greatest for Washington, DC and that it was lowest for Slough.

The location-dependent behaviour found by Smith and King (1981) is common in the ionospheric F-region. Modelling by Millward *et al.* (1996) and Zou *et al.* (2000) has shown that the variation of foF2 over the year at a given station is explained by changes to two key influences: i) thermospheric composition (which is influenced by a station's proximity to the geomagnetic pole) and ii) ion-production rate (which is influenced by solar zenith angle and the level of solar activity). The composition changes are related to other location-dependent effects, such as thermospheric winds, which blow F2 layer plasma up/down field lines where loss rates are lower/higher and this effect depends on the geomagnetic dip. For Slough, the annual variability in composition dominates over the zenith-angle effect resulting in the variation of foF2 being predominantly annual. However, at other locations, at similar geographic latitudes but different longitudes, a strong semi-annual variation is both observed and modelled, caused by the compositional changes between Equinox and Winter months being relatively small compared with the effect of the change in solar zenith angle. A method to determine and analyse the ratio of powers in the annual and semiannual variations has been presented by Scott, Stamper, and Rishbeth (2014) and used by Scott and Stamper (2015). We have extended this study to the Washington data and find, as for nearby stations studied by Scott and Stamper (2015), that the semi-annual variation dominates at Washington (and the variation of the annual/semiannual power ratio there is almost uncorrelated with that at Slough). Thus ionising solar EUV irradiance is more important in controlling foF2 at Washington than it is at Slough where the composition effect (on loss rates) dominates. EUV emission (particularly at the softer end of the spectrum) is enhanced through the presence around sunspots of plages and faculae (Dudok de Wit *et al.*, 2008) and hence foF2 is expected to be more dependent on both sunspot numbers and facular area at Washington than at Slough.

The results of Smith and King (1981) also help to explain the non-linearity of the foF2–$R$ variations that can be seen in Figure 2 (often called the saturation effect, see also Sethi, Goel, and Mahajan, 2002). This is because the RGO facular areas increase with $R$ at lower $R$ but reach a maximum and then fall again at the largest $R$ (Foukal, 1993). Of all the sites studied by Smith and King (1981), Slough had the lowest sensitivity to facular area. The Slough data also show the lowest solar-cycle hysteresis in the foF2–$R$ relationship for a given UT. Indeed, analysis by Bradley (1994) found that for Slough there were no detectable cycle-to-cycle changes in the average foF2 variations with $R$ (at a given UT) in that they were smaller than the solar-cycle hysteresis effect (which was not systematic) and both geophysical and observational noise.

## 2. Slough Ionosonde Data

Figure 1c shows the Slough ionosonde foF2 data, retrieved from the UK Space Science Data Centre (UKSSDC) at RAL Space, Chilton (URL given at the end of this section).  In 2004, a more complete set of scaled and tabulated hourly data for 1932–1943 was re-discovered in the archives of World Data Centre C1 at Chilton.  These data have been digitised and checked wherever comparisons are possible and by re-scaling a few selected ionograms from the surviving original photographic records.  A few soundings were not usable because the metadata revealed that the ionosonde was operated in a mode unsuitable for foF2 determination.  Regular soundings at Noon began in February 1932, and after January 1933 the sounder was operated six days a week until September 1943, when regular hourly soundings every day began. Before 1943 values for noon were available every day but for other UTs only monthly medians were tabulated (of a variable number of samples, but always exceeding 15).  Interference was not a problem for the earliest data as the HF radio spectrum was not heavily utilised, but some data carry a quality flag "C" that appears to stand for "cows", who caused a different kind of interference by breaking through the fence surrounding the neighbouring farm and disrupting performance by scratching themselves against the receiver aerials.  The hardware used (at least until later in the data series) was constructed in-house and evolved from the first sounder made by L.H. Bainbridge-Bell, to the 249 Pattern, the Union Radio Mark II, and the KEL IPS42.   In the present article, annual means of foF2 were compiled for each of the 24 Universal Times (UT) separately:  for regular hourly values a total of at least 280 soundings in a year (≈75 %) were required to make a useable annual mean, and for monthly median data ten values per year were required.  In Figure 1c it can be seen that the noise in the annual mean data is considerably greater before 1943 for most UTs. This could be due to the use of monthly medians rather than the monthly means of daily values and the fact that data were only recorded six days per week, but also potentially associated with the stability of the sounder and observer scaling practices.  However, the values for Noon (the black line) show the same year-to-year consistency before and after 1943, implying that the use of medians and the reduced sampling is the main cause of the increased noise in the earliest data.  The grey lines in Figure 1c are for UTs at which the correlation coefficient between $R$ and $R_{\text{fit}}$, the best third-order- polynomial fit of foF2 to $R$ for data after 1950 (see next section), does not exceed 0.99, whereas the coloured lines are for UTs (mainly during the daytime) for which this correlation does exceed 0.99.

After 1990, the ionosonde at Slough was relocated to Chilton, Oxfordshire. To avoid the need for a data intercalibration between these two sites, and any potential effects that may have, we here only consider data up to an end date of 1990.

## 3. Analysis

As discussed in the introduction (and shown by Figure 2), in general, the relationship between foF2 and $R$ varies from cycle to cycle and with location. Smith and King (1981) used linear and polynomial multiple-regression fits to show that for all stations the part of the variation not well explained by sunspot number varied with the area of white-light faculae [$A_f$] on the visible solar disk, as measured by RGO before 1976. The part of the variation that was found to be associated with $A_f$ varied with location and of the six stations that they studied, the facular effect was smallest for Slough and largest for Washington. There is a correlation between facular area and CaK plage area but this is not exact: in particular, the "rollover" in $A_f$ at the highest $R$ is not seen in the plage area (Foukal, 1993). Nevertheless, multiple regressions between annual means of Slough foF2 and a combination of $R$ and plage area have been made by Kuriyan, Muralidharan, and Sampath (1983). Because CaK plage area varies monotonically with $R$ for annual means (Foukal, 1993), it should be possible to fit sunspot numbers with a polynomial in these foF2 data alone. *Bradley* (1994) used a second-order polynomial, and we here use a third-order one (but, in fact, the derived term in (foF2)$^3$ is usually relatively small). On the other hand, the Washington data (Figure 2) demonstrate that there are locations where the solar-zenith-angle effect dominates over composition effects (and hence the semiannual variation dominates over the annual) and there is a greater dependence on facular area [$A_f$]. Hence for the general case, we define the fitted $R$ from foF2 data as:

$$R_{fit} = \alpha \, \text{foF2}^3 + \beta \, \text{foF2}^2 + \gamma \, \text{foF2} + \delta + \varepsilon \, A_f \qquad (4)$$

Fits were made using the Nelder–Mead search procedure to minimise the r.m.s. deviation of $R_{fit}$ from the sunspot number in question ($R$, $R_{BB}$ and $R_C$). Note that the analysis presented below in this **article** was repeated using a second-order polynomial ($\alpha = 0$) and a linear fit ($\alpha = \beta = 0$). The results were very similar in all three cases, the largest difference being that uncertainties are smallest using the full third-order polynomial because fit residuals were smaller and had a distribution that was closer to a Gaussian. In the remainder of this article we show the results for the third-order polynomial but the overall results for the lower-order polynomials will also be given.

As expected from the results of Smith and King (1961), we found that in some cases the facular term was needed, whereas in others it was not. Specifically, the fits for Washington were statistically poorer if the facular term was omitted and so it was necessary to use $\varepsilon \neq 0$. On the other hand, for Slough there was no statistically significant difference between the fits with ($\varepsilon \neq 0$) and without ($\varepsilon = 0$) the facular term: to demonstrate this we here discuss both the Washington and the Slough fits, both with and without the facular area. In sub-section (3–i) the fits employ a third-order polynomial in Slough foF2 only (i.e., $\varepsilon = 0$) whereas in section (3–ii) we fit the same data using the third-order polynomial in foF2 plus a linear term in the RGO white-light facular area, $A_f$, (i.e., $\varepsilon \neq 0$). The latter fits only use data before 1976, when the RGO measurements ceased. Both $\varepsilon = 0$ and $\varepsilon \neq 0$ fits can be carried out for the Sough data (and are shown to give similar results) because the dependence on $A_f$ is low. In sub-section (3–iii) we study the Washington data and find the greater dependence on facular area means this factor must be included. (Without the $\varepsilon A_f$ term, the correlations between $R_{fit}$ and sunspot numbers for Washington fall short of the required threshold that we here adopt). We note that fitted $\alpha$ values make the $\alpha.foF2^3$ term small and inclusion of the $\varepsilon A_f$ term makes the $\beta.foF2^2$ small also, such that $R_{fit}$ is approximately a combination of linear terms in foF2 and $A_f$, as was found by Smith and King (1981).

The sources of the data used in the following sections are the following: the Slough foF2 data and the Greenwich white-light facular area data were downloaded from the World Data Centre (WDC) for Solar Terrestrial Physics, which is part of the UK Space Science Data Centre (UKSSDC) at RAL Space, Chilton, UK (http://www.ukssdc.ac.uk/wdcc1/ionosondes/secure/iono_data.shtml); the Washington foF2 data were downloaded from Space Weather Services in Sydney, Australia (formerly known as IPS and the WDC for Solar-Terrestrial Science) within the Australian Bureau of Meteorology (ftp://ftp-out.ips.gov.au/wdc/iondata/medians/foF2/7125.00); the standard sunspot numbers [$R$] are the old data series published (until July 2015) by the WDC for the sunspot index part of the Solar Influences Data Analysis Center (SIDC) at Royal Observatory of Belgium (http://sidc.oma.be/silso/versionarchive). The corrected sunspot numbers series [$R_C$] is given in the supplementary data to the article by Lockwood *et al.* (2014) and the backbone sunspot group data [$R_{BB}$] were digitised from the article by Svalgaard and Schatten (2015) that accompanied the call for articles for this special issue. We employ the version of the RGO sunspot-group data made available by the Space Physics website of the Marshall Space Flight Center (MSFC) which has been compiled, maintained and corrected by D. Hathaway. These data were downloaded in June 2015 from http://solarscience.msfc.nasa.gov/greenwch.shtml. As noted by Willis *et al.* (2013b), there are some differences between these MSFC data and versions of the RGO data stored elsewhere (notably those in the National Geophysical Data Center, NGDC, Boulder, http://www.ngdc.noaa.gov/nndc/struts/results?op_0=eq&v_0=Greenwich&t=102827&s=40&d=8&d=470&d=9), but these are very minor.

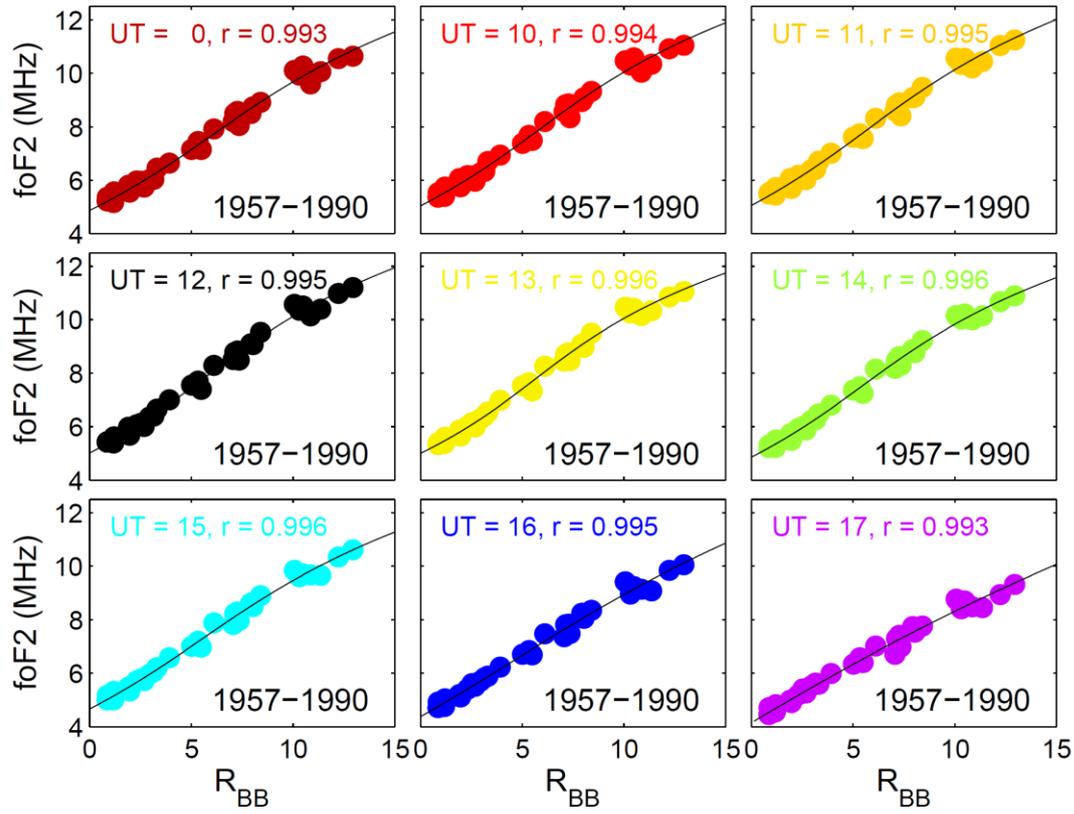

**Figure 3**. Slough foF2 values for 1957−1900 as a function of $R_{BB}$ for the nine UT's that yield a correlation coefficient r > 0.99 with all three of $R$, $R_{BB}$, and $R_C$. In each panel, points are observed 12-month running means and the lines are the best third-order-polynomial fits to which give fitted values $R_{fit}$ from the foF2(UT) values. The correlation coefficients [$r$] between $R_{fit}$ and $R_{BB}$ are given for each of these UT.

(3.1) **Using Slough Data and Polynomial Fits in foF2 Only (ε = 0)**

Figure 3 shows values of 12-month running means of foF2 at various UTs as a function of sunspot number for the interval 1957−1990. This calibration interval contains no information from the putative Waldmeier discontinuity and before. Plots were made for $R_{BB}$, $R_C$, and $R$ and the results are very similar in form and so only the results for $R_{BB}$ are shown here. The black lines are the best-fit to $R$ (minimum r.m.s. residual), third-order polynomial in foF2, $R_{fit}$. The selection of UTs shown is explained below. The values of $R_{fit}$ as a function of $R_{BB}$ are shown in Figure 4, with data points (open circles) coloured using the same colour scheme as Figure 3. The diagonal black line shows the ideal fit line: $R_{BB} = R_{fit}$. The colour key gives the correlation coefficients [$r$] between $R_{BB}$ and $R_{fit}$ for the different UTs. As for Figure 3, plots using $R$ or $R_C$ are almost identical to those for $R_{BB}$ and are not shown.

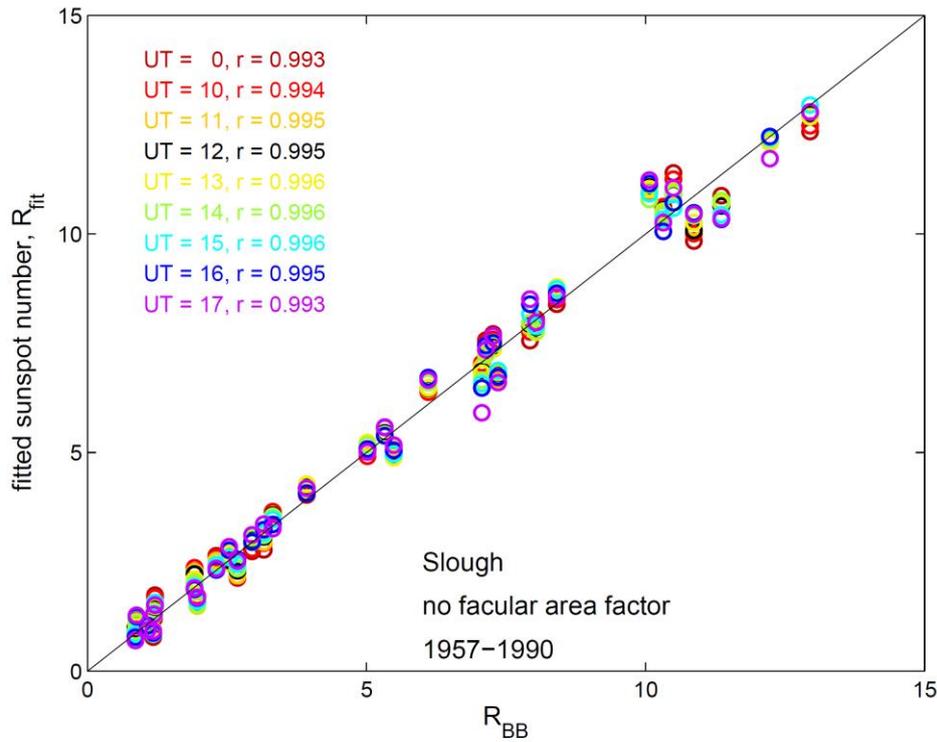

**Figure 4.** The values from the third-order-polynomial fits shown in Figure 3 [$R_{fit}$] as a function of $R_{BB}$. The solid line is the ideal fit of slope unity and intercept zero. The colours give the UT and correlation coefficient for that UT, as in Figure 3. As for Figure 3 these data are from the interval 1957−1990.

The variation of the correlation coefficient with UT for $R_{BB}$ is shown by the red line in Figure 5. The blue and green lines in Figure 5 show the results for $R$ and $R_C$, which are identical (because for this calibration interval of 1957−1990, $R$ and $R_C$ are identical). In Figures 3 and 4, only the UTs meeting the criterion that the correlation *r* exceeds 0.99 are used. This threshold selects nine UTs from the available twenty-four. Our study was repeated for lower thresholds (from the largest possible value of 0.996, which gives just one UT that meets the criterion, down to 0.98 at which all twenty-four UTs qualify) and it was found that the estimated 2σ uncertainties in the analysis discussed below were lowest for the threshold of 0.99.

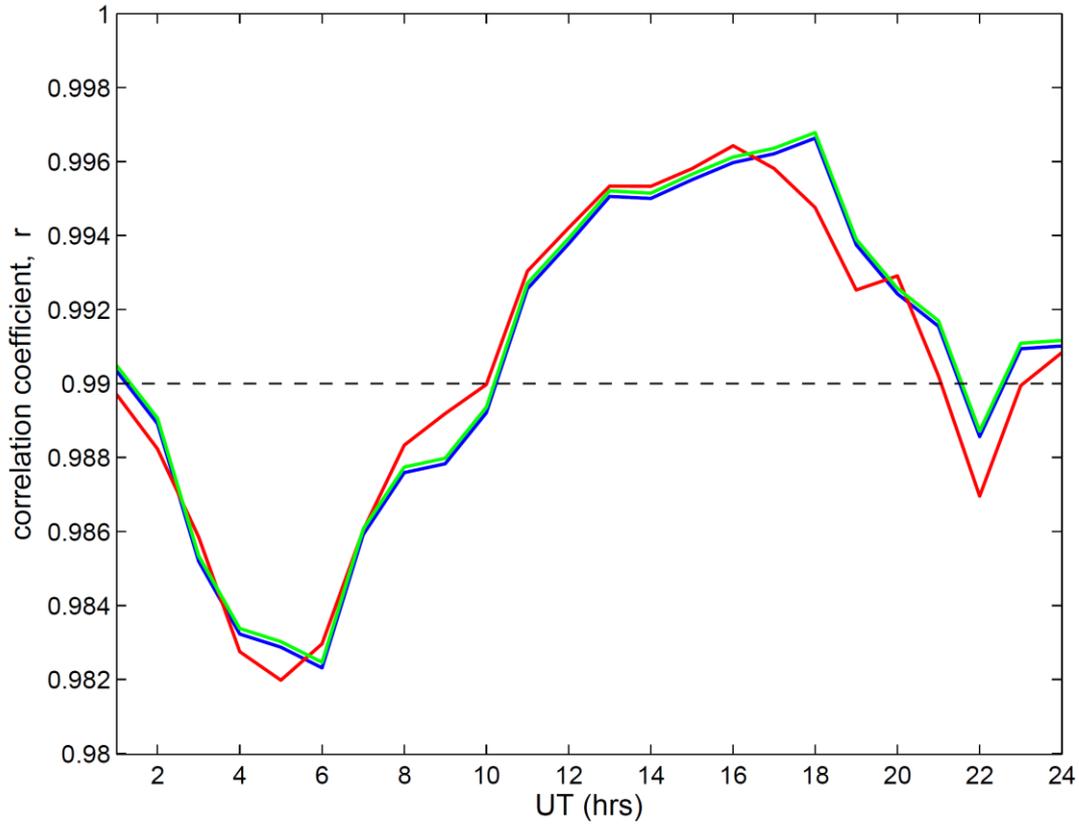

**Figure 5**. Correlation coefficients [$r$] between observed ($R$, $R_{BB}$, and $R_C$) and corresponding fitted sunspot numbers, $R_{fit}$, as a function of UT for the interval 1957-1990. The blue line is for $R$, the red line for $R_{BB}$ and the green line for $R_C$. (Note $R$ and $R_C$ are, by definition, the same for this interval).

Figure 6 shows the temporal variation of the mean fit residuals. These have been evaluated for all **of** the data (from the present back to 1932) and not just the calibration interval (1957–1990) used to derive the coefficients of the best-fit third-order polynomial ($\alpha$, $\beta$, $\gamma$**,** and $\delta$; remember that $\varepsilon$ is taken to be zero in this section). The means are calculated over all UTs for which the correlations $r$ between all three sunspot measures and their corresponding $R_{fit}$ values exceed 0.99 for the calibration interval. In order to display the longer-term trends in the sunspot calibrations, 11-year running means have been taken. The grey areas mark the 2$\sigma$ uncertainty band around the means (where $\sigma$ is the standard deviation).

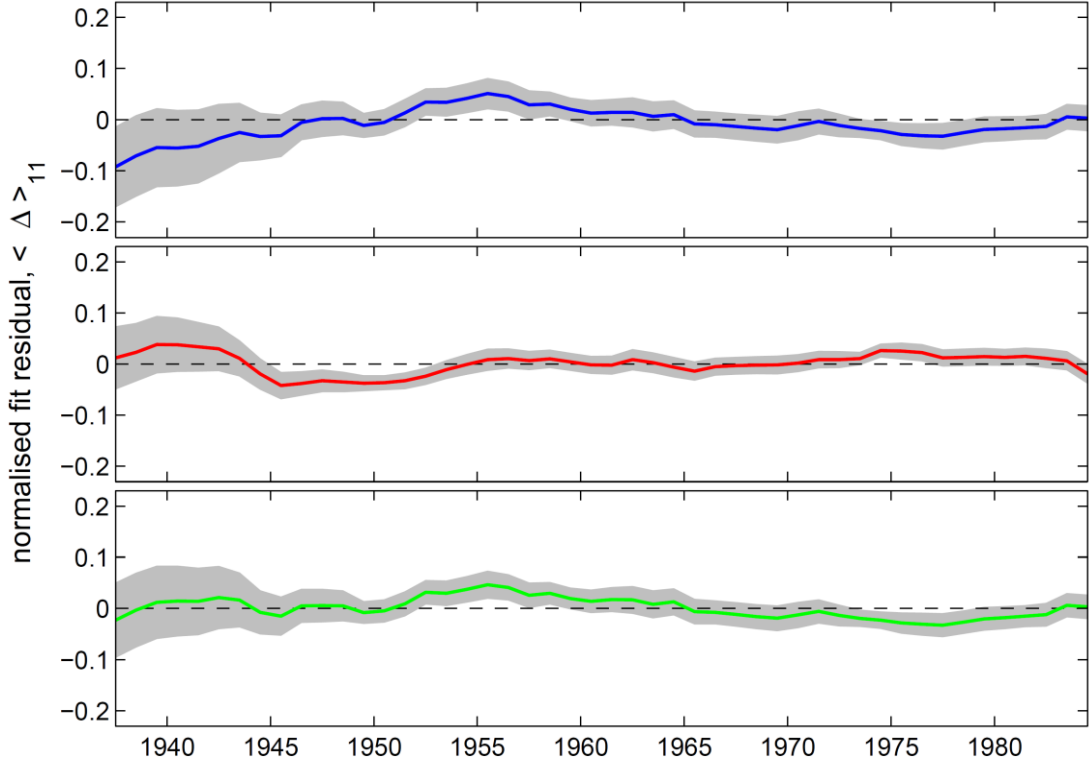

**Figure 6**. Mean normalised fit residuals as a function of time, for: (top, in blue) $R$; (middle, in red) $R_{BB}$; and (bottom, in green) $R_C$. In each case, the grey area marks the band between ±2σ around the mean values. 11-year running means of annual values are shown to highlight long-term trends.

In the top panel, the blue line shows that for the standard sunspot number [$R$] the fit residuals are small and reasonably constant for the calibration interval (and after 1990) but become persistently negative before then. This means that $R$ in this interval is systematically smaller than the best-fit extrapolation based on foF2. The deviation is slightly smaller than the 2σ uncertainty (but exceeds the 1σ uncertainty). The sense of this persistent deviation is consistent with the Waldmeier discontinuity. The second panel is for $R_{BB}$. In this case, the red line shows even better fits during the calibration interval and after, but $R_{BB}$ for before 1945 becomes consistently greater than the best-fit extrapolation from the calibration interval. Again the deviation is slightly smaller than the 2σ uncertainty but is almost as large in magnitude as that for $R$. Thus the Slough foF2 data imply that the effective correction for the Waldmeier discontinuity in $R_{BB}$ is roughly twice what it should be for the 20% correction postulated by Svalgaard (2011), as was found by Lockwood, Owens, and Barnard (2014). In the third panel, the green line shows the results for $R_C$ which uses the 11.6% best-fit correction found by Lockwood, Owens, and Barnard (2014). In this case the fit residuals before the calibration interval are similar to those during it.

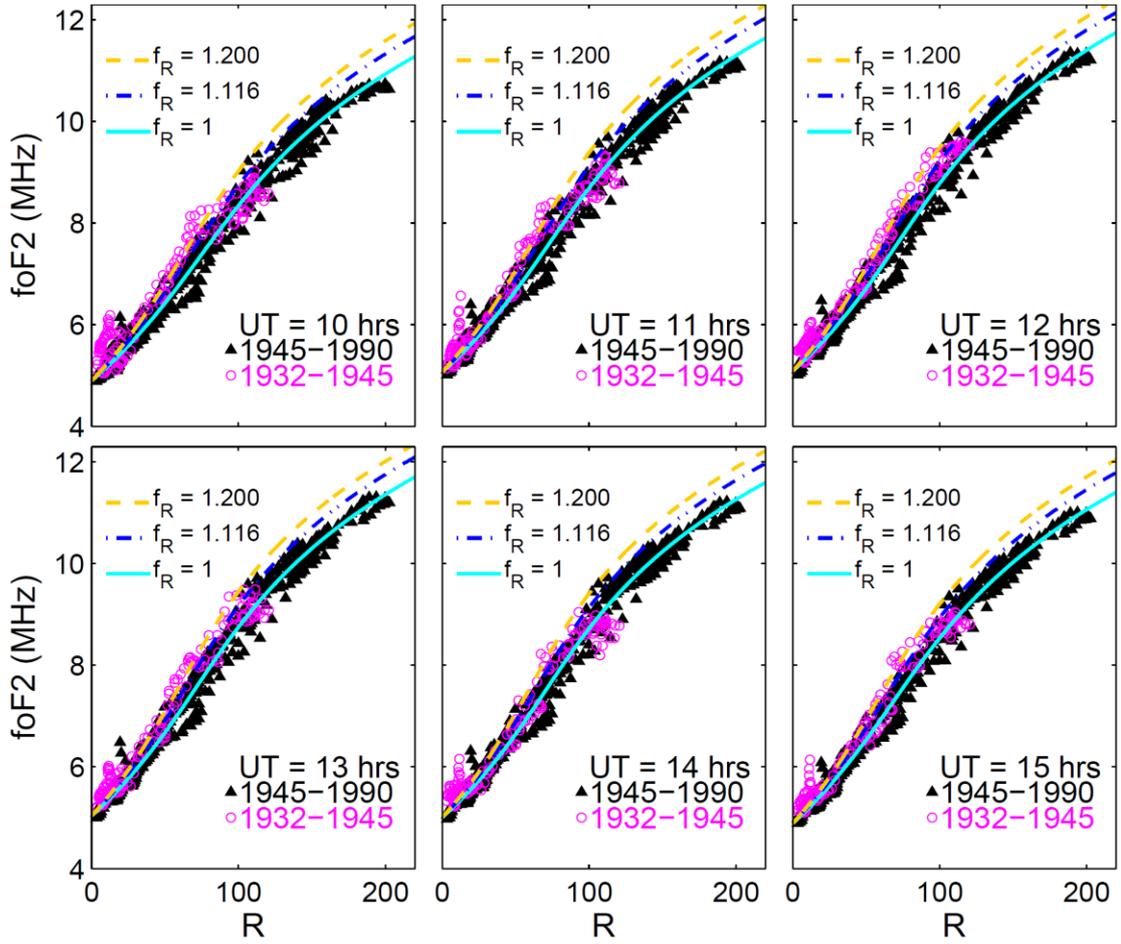

**Figure 7**. Scatter plots of annual means of foF2 as a function of version 1 of the international sunspot number $R$ for the six UTs for which $r \geq 0.99$ for both the "before" (mauve open circles) and "calibration" (black solid triangles) intervals (1932−1945 and 1945−1990, respectively). The solid cyan line is the best-fit third-order polynomial to the calibration data. The dot-and-dash blue and dashed orange lines are for this best-fit $R$ divided by $f_R$ of 1.116 (derived by Lockwood, Owens, and Barnard, 2014) and 1.200 (derived by Svalgaard, 2011) on which the mauve ("before") points should all lie if the correction needed for the Waldmeier discontinuity were 11.6% and 20%, respectively.

Here we also use the annual means of the Slough data to see if they agree with the correction factor derived from the RGO sunspot-group area and number by Lockwood, Owens, and Barnard (2014). The analysis was carried out for a "before" interval of 1932–1945 and "after" intervals of 1945–1990 (which uses all of the data and gives the results shown in Figure 7) and 1945–1959 (which makes the "before" and "after" intervals of equal length) and 1957–1990 (over which $R_{BB}$, $R_C$, and $R$ agree most closely). The results were essentially the same and conclusions drawn do not depend on the intervals adopted. There are six UTs for which the correlation $r$ between $R$ and its $R_{fit}$ variation exceeds the 0.99 criterion for both the 1932−1945 interval (before the putative Waldmeier discontinuity) and for the 1957–1990 calibration interval (after the putative Waldmeier discontinuity). Scatter plots of foF2 for these six UTs are shown in Figure 7. In each case black filled triangles are for the "after" (calibration) interval and the mauve open circles are for the "before" interval. It is significant that the behaviour for 12 UT is the same as for the other UTs that meet the criterion (which are all for daytime UTs) because the ionospheric test data are based on monthly means of hourly data in this case, whereas for all other UTs they are based on monthly medians. However this difference has not had any discernible effect.

As implied by Figure 6, the values of $R$ before 1945 are, on average, slightly lower than those in the "after" interval at the same foF2. Otherwise the variations of foF2 with $R$ for the two intervals are very similar. The solid cyan line in each panel is the best-fit third order polynomial to the calibration data. It can be seen that most mauve points lie above the cyan line, consistent with R being underestimated before the Waldmeier discontinuity. The blue dot-and-dash line and the dashed orange line are for this best-fit $R$ divided by correction factors $f_R$ of 1.116 and 1.200 on which the mauve points should (if the real $R$–foF2 variation has remained the same) all lie if the correction needed for the Waldmeier discontinuity were 11.6 % (as derived by Lockwood, Owens, and Barnard, 2014) or 20 % (as derived by Svalgaard, 2011), respectively. The separations of the lines are small but inspection shows that the "before" interval test points (in mauve) are most clustered around the blue dashed lines (51.5 % of all the mauve points in all the panels in Figure 7 line lie below the blue dashed lines whereas 49. 5 % lie above it). In contrast, 73 % of all the points lie below the orange lines and only 27 % above, strongly implying that $f_R = 1.200$ is an overestimate of the correction needed.

To quantify the $f_R$ that is implied by the ionosonde data with greater precision, Figure 8 applies the procedure used by Lockwood, Owens, and Barnard (2014) to these Slough ionosonde data. The Waldmeier discontinuity is taken to be at 1945 and to be such that $R$ values before this time should be $f_R R$, instead of the standard $R$ value. The factor $f_R$ was varied between 0.9 and 1.3 in steps of 0.001. The difference between mean fit residuals for "before" and "after" (calibration) intervals [respectively $<\delta R>_b$ and $<\delta R>_a$] was evaluated as a function of the factor $f_R$, where $\delta R = R_{fit} - R$. These means include the fit residuals for all six UTs for which both "before" and "after" intervals meet the r ≥ 0.99 criterion. As shown in 8(a), as $f_R$ is increased, $<\delta R>_b - <\delta R>_a$ falls linearly (because $<\delta R>_b$ is reduced as $f_R$, and hence $R$ in the "before" interval, is increased) and the ideal correction factor is when $<\delta R>_b - <\delta R>_a = 0$ as this means there is no longer a systematic offset between the "before" and "after" intervals, relative to the test data.

The probability p-value for each difference between the two means is computed using the procedure described by Lockwood, Owens, and Barnard (2014) and in Section 1.2. This peaks when the difference falls to zero but also gives the probability for all other values of $f_R$. All p-value distributions are normalised so that the area under the curve is unity.

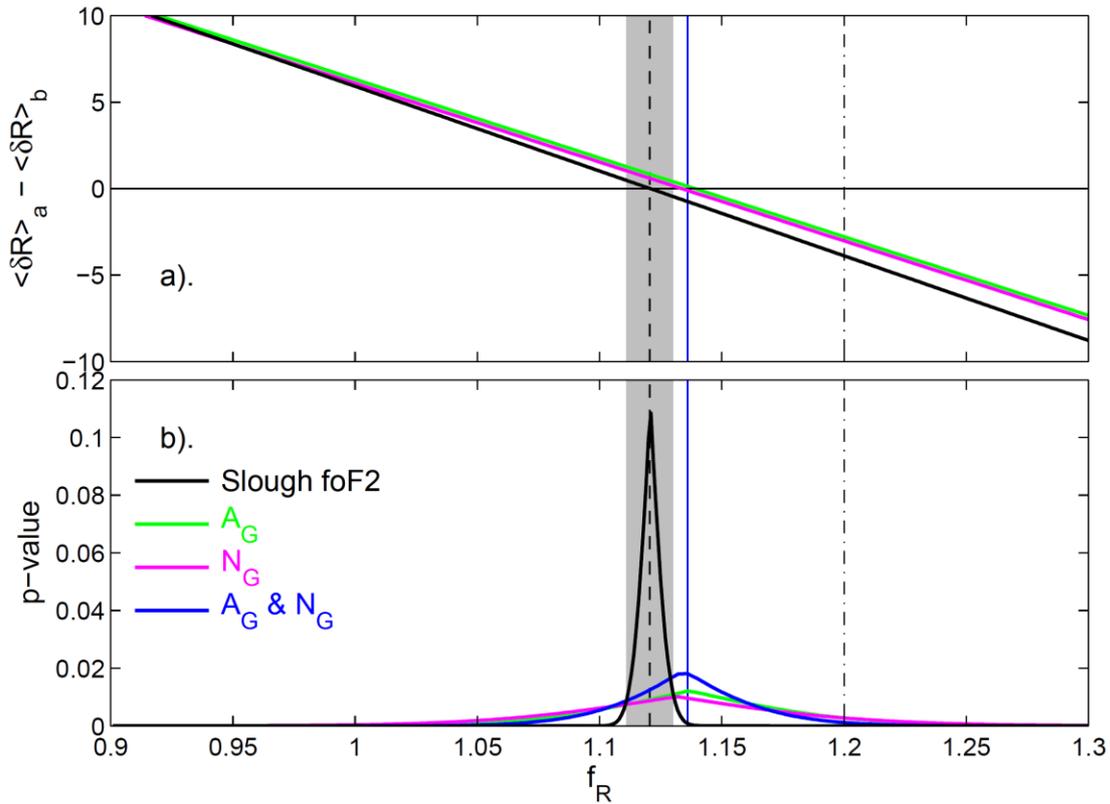

**Figure 8**. (a) The difference between mean fit residuals for "before" and "after" intervals [respectively, $<\delta R>_b − <\delta R>_a$] as a function of the correction factor $f_R$ applied to the before interval. (b) the p-value for that difference (see text). The green and mauve lines are the same for the RGO sunspot group area and number ($A_G$ and $N_G$ respectively), as used by Lockwood, Owens, and Barnard (2014) but here applied to a much shorter "before" interval of 1932−1945 and an "after" interval of 1945−1976. The blue line is the p-value for the combination of $A_G$ and $N_G$. The black line is the value for the six UTs for which the correlation coefficient between $R$ and $R_{fit}$ (using foF2 only, i.e. $\varepsilon = 0$), $r$, exceeds 0.99 for both the before and after interval. The vertical solid blue line is the optimum $f_R$ from the combination of $A_G$ and $N_G$ and the vertical dashed line is the peak value derived from the Slough foF2 data. The dot-and-dash line is the value of $f_R$ proposed by Svalgaard (2011) that is inherent in the $R_{BB}$ data series. The grey area shows the $2\sigma$ uncertainty band for the fits using foF2.

In addition to carrying out this test using the Slough foF2 data, we have repeated it for RGO sunspot group number [$N_G$] and the RGO sunspot group area [$A_G$]. This test is the same as that which was carried out by Lockwood, Owens, and Bernard (2014), except that here we use shorter intervals; the "before" interval being 1932–1945 (the same as for the foF2 data used here) and the "after" interval being 1945–1976. (Data between 1976 and 1990 were not used as they come from the SOON network and would require intercalibration with the RGO data). This eliminates any possibility that either drift in the early RGO data (before 1932) or the RGO–SOON calibrations are influencing our estimate of the optimum $f_R$.

The results are shown in Figure 8. In both parts of the figure, the black line is for the foF2 data, the green line is for $A_G$ and the mauve line is for $N_G$. The blue line is the combination of the $A_G$ and $N_G$ probability variations. The distribution for foF2 is very much narrower than those for $A_G$ and $N_G$ (meaning that the optimum value is much better constrained) and the peak p-value is therefore much greater. The vertical dashed line marks the peak for the foF2 test at $f_R$ =1.121 (*i.e.* a 12.1% correction) and the grey band marks the uncertainty band of ±2σ of the p-value distribution (between 1.1110 and 1.1298, *i.e.* a correction of 11.10–12.98%). This result was obtained employing a third-order-polynomial fit to the Slough foF2 data: if a second-order polynomial was used, the optimum value was 12.6% with a ±2σ uncertainty range of 11.11–14.17% and hence the optimum value is slightly higher and the uncertainty band considerably wider. To within the uncertainties, use of the second- and third-order polynomials gives the same result. If a linear variation was used, the optimum value was 13.85% with a ±2σ uncertainty range of 12.33–15.38%, which is a significantly higher value and with an uncertainty band that does not overlap with that for the third-order-polynomial analysis: however, this value is here discounted because the linear variation cannot reproduce the marked "rollover" in the foF2–R plots presented in Figure 3 and 7. The solid blue vertical line marks the optimum value from the combination of the $A_G$ and $N_G$ p-value distributions (at $f_R$ = 1.1360, *i.e.* a 13.60% correction) for the same intervals. This is slightly higher than the 11.6±3.3% correction found by Lockwood, Owens, and Bernard (2014), using the same test but applied to RGO data that extended back to 1874. This shows that the early RGO data had reduced the optimum correction factor derived from the RGO data somewhat, but only by 2%. This difference is comfortably within the ±3.35% uncertainty band estimate by Lockwood, Owens, and Bernard (2014). The dotted line is the 20% correction proposed by Svalgaard (2011), which is also inherent in $R_{BB}$. Because the p-value distributions for $N_G$ and $A_G$ are broad, the correction factor for $R$ of 12.0±1.0% derived here using foF2 is consistent with them, but using foF2 provides a much better-defined test value than the RGO sunspot data. The reason why the foF2 test constrains the required correction to a much greater extent than do the RGO data is two-fold: firstly the correlations for both the "before" and "after" intervals are so high ( ≥ 0.99); and secondly, the use of the six UTs that met this criteria means there is six times the number of datapoints available per year compared to either $N_G$ or $A_G$. The probability p-value from the foF2 test for a 20% correction is less than $10^{-20}$.

(3.2) **Using Slough Data with Polynomial Fits in foF2 and a Linear Dependence on Facular Area**

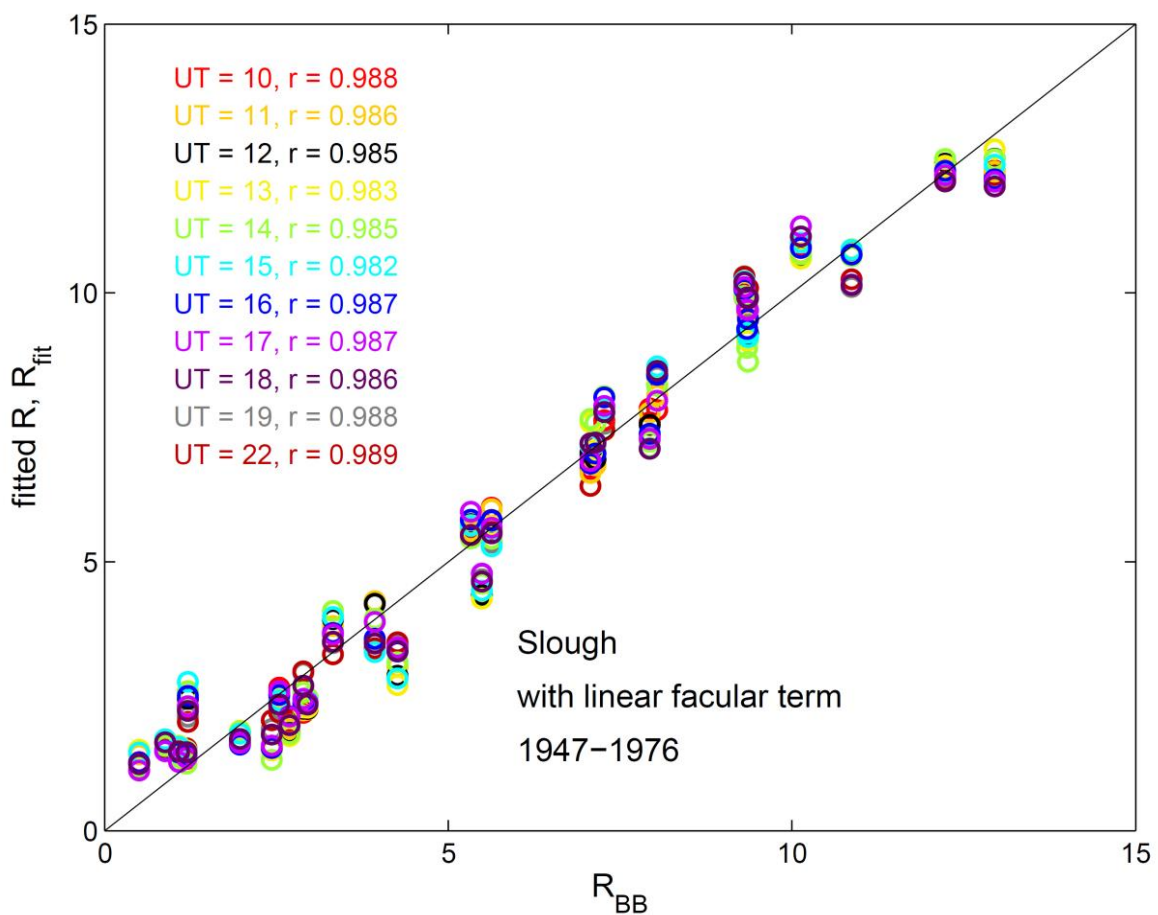

**Figure 9**. The same as Figure 4, but including a linear term for the white-light-facular area in the multivariate fit to the Slough foF2 data. Specifically, the values from the third-order polynomial fits [$R_{fit}$] are shown as a function of $R_{BB}$. The solid line is the ideal fit of slope unity and intercept zero. The colours give the UT and correlation coefficient for that UT. These data are from the interval 1947–1976.

As shown by Figure 2, the test presented in Section 3.1 will not work at all ionosonde stations and, in particular, those where Smith and King (1981) found a greater dependence of the R−foF2 relationship on facular area [$A_f$]. To test that this factor has not altered the results for the Slough data, we here repeat the analysis in Section 3.1 using a multivariate fit with a third-order polynomial in foF2 and a linear term in the RGO white light facular area [$A_f$] (*i.e.* $\varepsilon$ in Equation (4) is no longer assumed to be zero). Because RGO white light facular measurements ceased in 1976 we here use 1945–1976 for the "after" calibration interval. Otherwise the test is conducted as in the last section. The facular area observations were made at RGO and ceased in 1976. To maximise the number of data points after the Waldmeier discontinuity the interval 1947–1976 is used here.

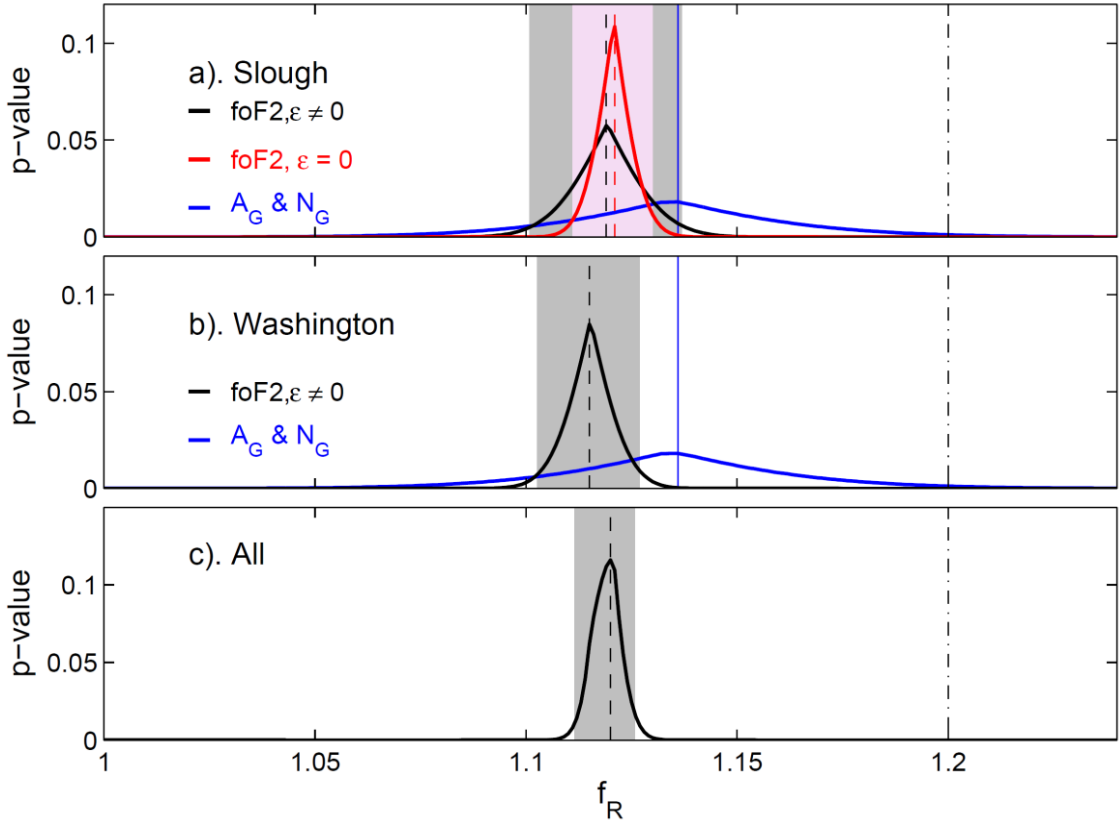

**Figure 10**. Plots of p-values of the difference between mean fit residuals for before and after intervals [$<\delta R>_b - <\delta R>_a$], as a function of the correction factor $f_R$ applied to the "before" interval. In Panel a, the red line is for the fit to Slough foF2 data with $\varepsilon = 0$ (as was presented in Figure 8b). The black line is also for fits to the Slough foF2 data but with $\varepsilon \neq 0$. The blue line in (a) and (b) is the distribution for RSO group areas and numbers ($A_G$ and $N_G$) combined. (b) is the same as (a) but the black line is for fits to the Washington foF2 data with $\varepsilon \neq 0$. (c) The total probability, the product of the p-values for four independent fits: with the RGO group number; the RGO group area; the Slough foF2 data with $\varepsilon = 0$; and the Washington foF2 data with $\varepsilon \neq 0$. The grey/pink bands mark the $2\sigma$ uncertainty band around the peaks of the black/red lines, respectively.

Figure 9 is the same as Figure 4, but this time including the $\varepsilon A_f$ term in $R_{fit}$. The agreement between $R_{BB}$ and $R_{fit}$ is again very good but no higher than in the last section, despite the additional fit parameter used in the fit. The additional noise introduced by the $A_f$ data mean that there are no UTs for which the correlation coefficient [$r$] between $R$ and $R_{\text{fit}}$, exceeds 0.99, however there are eleven for which $r$ exceeds 0.98, and these are shown in figure 9. The top panel of Figure 10 is the same format as Figure 8b, and the p-value distributions for the combination of $A_G$ and $N_G$ is shown in blue. This panel also repeats (in red) the distribution for Slough foF2 (with $\varepsilon = 0$) that was shown in Figure 8b (with the $\pm 2\sigma$ uncertainty band around the peak in pink). The black line is the corresponding distribution for the Slough data with the best fit $\varepsilon$, inherent in the fits shown in Figure 9. The main effect is that the p-value distribution is slightly broadened when the facular term is introduced (the $2\sigma$ uncertainty shown in grey is raised from $\pm 1.0\%$ to $1.8\%$), but the peak value is hardly altered (11.9% instead of 12.1%). The additional uncertainty in the optimum value of $f_R$ is associated with the additional noise introduced by the facular area data. Hence for this test using the Slough foF2 data, the main effect of allowing for the facular area is to increase the noise level.

## (3.3) Using Washington Data with Polynomial Fits in foF2 and a Linear Dependence on Facular Area

From the previous two sections, we find that in the case of the Slough data, adding the linear term in facular area does not make a significant difference to the best estimate of the correction factor. This is not true of the Washington data, for which Smith and King (1981) found a greater dependence on facular area. Figure 11 is the same as Figure 9 for the Washington data and shows that with allowance for the facular area effect, a good fit can be obtained. The middle panel of Figure 10 shows the p-value distribution derived from these fits in black, and the optimum correction factor is 11.5±1.2%. Note that, unlike for the Slough data, the test for $\varepsilon = 0$ cannot be carried out for the Washington data to the level of accuracy we require as no UT in the "after" interval meets the requirement that the correlation $r$ between $R$ and $R_{\text{fit}}$ exceed 0.99.

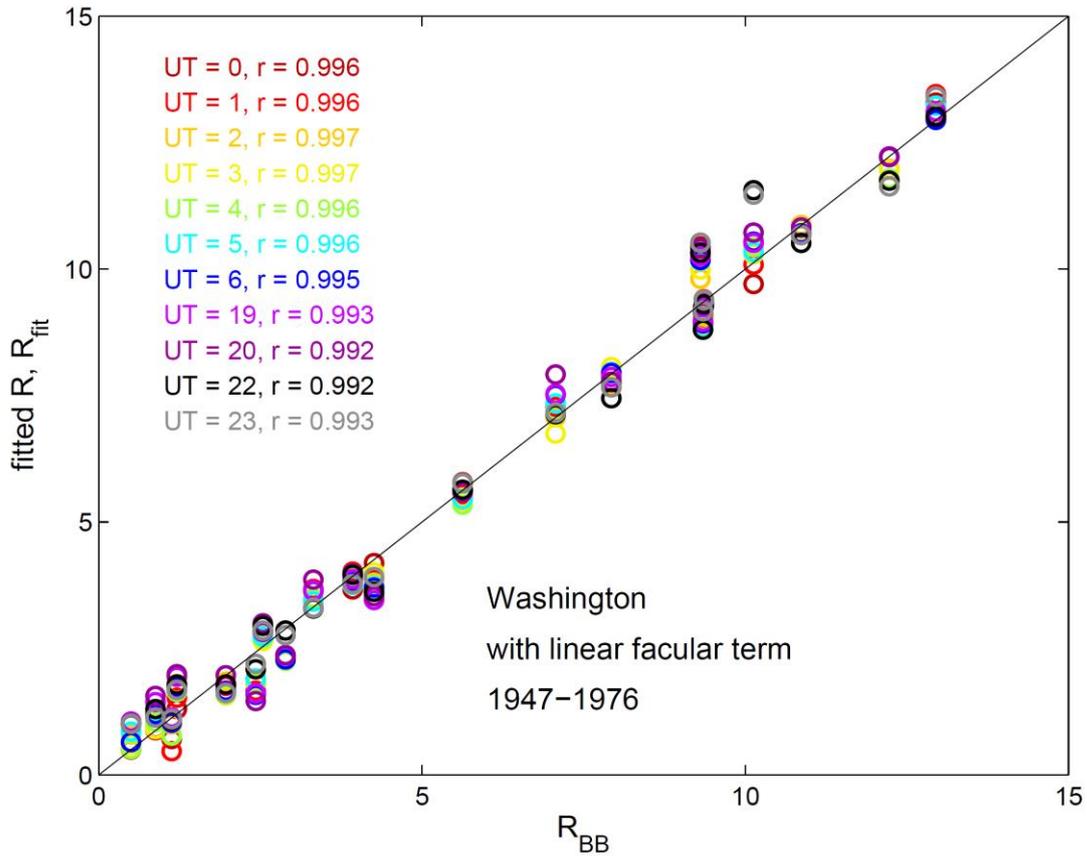

**Figure 11**. The same as Figure 9 for Washington foF2 data. Specifically, the values from the third-order polynomial fits with a linear term in the white-light facular area [$R_{fit}$] are shown as a function of $R_{BB}$. The solid line is the ideal fit of slope unity and intercept zero. The colours give the UT and correlation coefficient for that UT. As for Figure 9, these data are from the interval 1947−1976.

(**3.4**) **Comparing and Combining Slough and Washington Ionospheric Data and RGO Data**

Panels a and b of Figure 11 show that the $R_{fit}$ values from foF2 data imply very similar $f_R$ factors (which need to be applied to $R$ to allow for the Waldmeier discontinuity) as are derived from the RGO sunspot data. We combine the results of all the data that give $r > 0.99$ from the last three sections by multiplying independent p-value distributions together. The black line in the bottom panel shows the product of results for: Slough foF2 with $\varepsilon = 0$; Washington foF2 with $\varepsilon \neq 0$; and RGO $N_G$ and RGO $A_G$. Again the grey band is the uncertainty around the peak at the ±2σ level. If we make the assumption that the differences between "before" and "after" ionospheric data are due only to changes in the calibration of $R$, Figure 11c shows that all tests give an optimum $f_R$ from the combination of all the independent p-value distributions of 12.0 % (the peak of the black line in Figure 11c) and the 2σ uncertainty band around this optimum value (in grey in Figure 11c) is 11.16-12.57 %.

## 4. Conclusions

We conclude that the ionosonde data give an extremely accurate test for the Waldmeier discontinuity correction factor and the best value (maximum p-value with a $\pm 2\sigma$ uncertainty) from a combination of the Slough and Washington ionospheric data is $12.0\pm1\%$, which is very similar to the results obtained from the RGO sunspot group data. In general, allowance for the dependence of the foF2–R relationship on the facular area [$A_f$] is required but is sufficiently small for Slough (where foF2 is dominated by the composition effect) that the results are essentially the same if it is neglected. For Washington (where foF2 is dominated by the solar illumination effect), the $A_f$ factor cannot be neglected. The probability that the Waldmeier correction is as large as the $\approx 20\%$ adopted by Svalgaard (2011) and the >20% that is inherent in the backbone sunspot number $R_{BB}$ is, by this test, essentially zero. The results show that $R_{BB}$ is $12.0\pm1.0\%$ too large for 1945 and before. The dashed red line in the middle panel of Figure 1 shows that the effect of applying this correction to $R_{BB}$ makes it almost identical to $R_C$ for the interval studied here.

The fact that $R_{BB}$ matches the best-fit ionospheric data better than the other series after the Waldmeier discontinuity reveals a very important implication. This improvement is because $R_{BB}$ corrects for a drift in the $k$-values for the Locarno Wolf numbers (Clette *et al.*, 2015). This drift was found by research aimed at explaining why the relationship between the F10.7 radio flux and international sunspot number (Johnson, 2011) broke down so dramatically just after the long and low activity minimum between Cycles 23 and 24. The Locarno $k$-values were re-assessed using the average of sixteen other stations (out of a total of about eighty) that provided near-continuous data over the 32-year interval studied. The results showed that the Locarno $k$-factors had varied by between +15% in 1987 and −15% in 2009. Before these tests were done the Locarno $k$-values formed the "backbone" of the international sunspot number series and were assumed to be constant. Note that this drift of 30% occurred, and went undetected, in this key backbone for twenty-two years in the modern era, despite there being at least eighty observatories available, and with defined and agreed procedures and related test data available such as F10.7. We have to be aware that in earlier times, with fewer stations, less well-defined procedures, less stable instrumentation, and with fewer (if any) data to check against, larger drifts will almost certainly have occurred in the prior "backbone" data series that are daisy-chained to generate $R_{BB}$.

Using ionosonde data we can only test the sunspot-number series back to 1932. But even at this relatively late date, the tests using the Slough and Washington ionosonde data indicate that $R_{BB}$ is significantly too large. Given the daisy-chaining of intercalibrations involved in the construction of $R_{BB}$, all values before 1945 need to be 12% smaller (relative to modern values) to make proper allowance for the Waldmeier discontinuity. However, the difference between $R$ (or $R_C$) and $R_{BB}$ also grows increasingly large as one goes back in time (see Article 2, Lockwood *et al.*, 2015a): from the study presented here we cannot tell if this trend has the same origin as the detected difference during Cycle 17; however Cycle 17 is consistent with the longer-term trend. That an error as large as the 12% can be found in $R_{BB}$ as late as 1945 does not give confidence that there are not much larger errors in $R_{BB}$ at earlier times.

**Acknowledgements** The authors wish to thank the staff of a number of data centres: the Slough foF2 data and the Greenwich white-light facular area data were downloaded from the World Data Centre (WDC) for Solar Terrestrial Physics, which is part of the UK Space Science Data Centre (UKSSDC) at RAL Space, Chilton, UK; the Washington foF2 data were downloaded from Space Weather Services in Sydney, Australia, part of the Australian Bureau of Meteorology; the international sunspot numbers (version 1) were downloaded (after July 2015 and from the archive section) from the WDC for the sunspot index part of the Solar Influences Data Analysis Center (SIDC) at the Royal Observatory of Belgium. The RGO sunspot group data made available by the Space Physics website of the Marshall Space Flight Center (MSFC). The work of M. Lockwood, C.J. Scott, M.J. Owens and L.A. Barnard at Reading was funded by STFC consolidated grant number ST/M000885/1.

**Disclosure of Potential Conflicts of Interest**

The authors declare that they have no conflicts of interest

**References**

Allen, C.W.: 1948, Critical frequencies, sunspots, and the Sun's ultra-violet radiation, *Terr. Magn. Atmos. Electr*. **53**, 433. doi:10.1029/TE053i004p00433.

Altschuler, M.A., Newkirk, G., Jr.: 1969, Magnetic fields and the structure of the solar corona, *Solar Phys*. **9**, 131. doi: 10.1007/BF00145734

Aparicio, A.J.P., Vaquero, J.M., Gallego, M.C.: 2012, The proposed "Waldmeier discontinuity": How does it affect to sunspot cycle characteristics?, *J. Space Weather Space Clim.* **2**, UNSP A12. doi:10.1051/swsc/2012012.

Bradley, P.A.: 1994, Further study of foF2 and M(3000)F2 in different **S**olar **C**ycles, *Annal.Geophys.* **37**, 2. doi: 10.4401/ag-4227

Chernosky, E.J., Hagan, M.P.: 1958, The Zurich sunspot number and its variations for 1700–1957, *J. Geophys. Res.* **63** 775. doi:10.1029/JZ063i004p00775.

Clette, F., Svalgaard, L., Vaquero, J.M., Cliver, E.W.: 2015, Revisiting the sunspot number, In *The Solar Activity Cycle*. eds. A. Balogh, H. Hudson, K. Petrovay and R. von Steiger, 35, Springer, New York. doi: 10.1007/978-1-4939-2584-1_3

Cliver, E.W., Ling, A.: 2015, The Discontinuity in ~1885 in the Group Sunspot Number, *Solar Phys.* (submitted)

Cliver, E.W., Clette, F., Svalgaard, L.: 2013, Recalibrating the Sunspot Number (SSN): The SSN Workshops, *Cent. Eur. Astrophys. Bull*. **37**, 401. ISSN 1845–8319,

Cnossen, I., Richmond, A.D.: 2008, Modelling the effects of changes in the Earth's magnetic field from 1957 to 1997 on the ionospheric hmF2 and foF2 parameters, J. Atmos. Sol. Terr Phys. **70**, 1512. doi: 10.1016/j.jastp.2008.05.003

Dudok de Wit, T., Kretzschmar, M., Aboudarham, J., Amblard, P.-O., Auchère , F., Lilensten, J.: 2008, Which solar EUV indices are best for reconstructing the solar EUV irradiance? *Adv. Space Res*. **42**, 903. doi: 10.1016/j.asr.2007.04.019

Foukal, P.: 1993, The curious case of the Greenwich faculae, *Solar Phys*. **148**, 219. doi: 10.1007/BF00645087

Foukal, P.: 2013, An explanation of the differences between the sunspot area scales of the Royal Greenwich and Mt.Wilson observatories, and the SOON program, *Solar Phys.* **289**, 1517. doi: 10.1007/s11207-013-0425-2

Gazis, P.R.: 1996, Solar cycle variation of the heliosphere, *Rev. Geophys.* **34**, 379. doi: 10.1029/96RG00892

Hoyt, D.V., Schatten, K.H.: 1994, The one hundredth year of Rudolf Wolf's death: Do we have the correct reconstruction of solar activity?, *Geophys. Res. Lett.* **21**, 2067. doi:10.1029/94GL01698

Hoyt, D.V., Schatten, K.H.: 1998, Group sunspot numbers: A new solar activity reconstruction, *Solar Phys.* **181**, 491. doi:10.1023/A:1005056326158

Ikubanni, S.O., Adebesin, B.O., Adebiyi, S.J., Adeniyi, J.O.: 2013, Relationship between F2 layer critical frequency and solar activity indices during different solar epochs, *Indian J. Radio Space Phys.* **42**, 73.

Johnson, R.W.: 2011, Power law relating 10.7 cm flux to sunspot number, *Astrophys. Space Sci.* **332**, 73. doi: 10.1007/s10509-010-0500-1

Krivova, N.A., Balmaceda, L., Solanki, S.K.: 2007, Reconstruction of solar total irradiance since 1700 from the surface magnetic flux, *Astron. and Astrophys.* **467**, 335, doi: 10.1051/0004-6361:20066725

Krivova, N.A., Solanki, S.K., Wenzler, T., Podlipnik, B.: 2009, Reconstruction of solar UV irradiance since 1974, *J. Geophys. Res.* **114**, D00I04. doi:10.1029/2009JD012375

Kuriyan, P.P., Muralidharan, V., Sampath, S.: 1983, Long-term relationships between sunspots, Ca-plages and the ionosphere, *J. Atmos. Terr. Phys*. **45**, 285. doi: 10.1016/S0021-9169(83)80034-8

Lockwood, M., Owens, M.J.: 2014a, Centennial variations in sunspot number, open solar flux and streamer belt width: 3. Modelling, *J. Geophys. Res. (Space Phys.)* **119**, 5193. doi: 10.1002/2014JA019973

Lockwood, M., Owens, M.J.: 2014b, Implications of the recent low solar minimum for the solar wind during the Maunder minimum, *Astrophys. J. Lett.* **781**: L7. doi:10.1088/2041-8205/781/1/L7

Lockwood, M., Owens, M.J., Barnard, L.: 2014, Centennial variations in sunspot number, open solar flux, and streamer belt width: 1. Correction of the sunspot number record since 1874, *J. Geophys. Res. (Space Phys.)* **119**, 5193.  doi:10.1002/2014JA019970.

Lockwood, M., Owens, M.J., Barnard, L.A.: 2015, Tests of sunspot number sequences: 4. Discontinuities around 1945 in various sunspot number and sunspot group number reconstructions, *Solar Phys.* (submitted).

Lockwood, M., Rouillard, A., Finch, I., Stamper, R.: 2006, Comment on "The IDV index: Its derivation and use in inferring long-term variations of the interplanetary magnetic field strength" by Leif Svalgaard and Edward W. Cliver, *J. Geophys. Res. (Space Phys.)* **111**, A09109. doi:10.1029/2006JA011640.

Lockwood, M., Nevanlinna, H., Barnard, L., Owens, M.J., Harrison, R.G., Rouillard, A.P., Scott, C.J.: 2014, Reconstruction of Geomagnetic Activity and Near-Earth Interplanetary Conditions over the Past 167 Years: 4. Near-Earth Solar Wind Speed, IMF, and Open Solar Flux, *Annales. Geophys*. **32**, 383. doi:10.5194/angeo-32-383-2014

Lockwood, M., Scott, C.J., Owens, M.J., Barnard, L., Nevanlinna, H.: 2016a, Tests of sunspot number sequences. 2. Using geomagnetic and auroral data, *Solar  Phys.* (submitted).

Lockwood, M., Owens, M.J., Barnard, L., Usoskin, I.G.: 2016b, Tests of sunspot number sequences. 3. Effects of regression procedures on the calibration of historic sunspot data, *Solar Phys*., in press, 10.1007/s11207-015-0829-2

Özgüç, A., Ataç, T., Pektaş, R.: 2008, Examination of the solar cycle variation of foF2 for cycles 22 and 23, *J. Atmos. Sol.-Terr. Phys*. **70**, 268. doi:10.1016/j.jastp.2007.08.016

Owens, M.J., Lockwood, M.: 2012, Cyclic loss of open solar flux since 1868: The link to heliospheric current sheet tilt and implications for the Maunder Minimum, *J. Geophys. Res*. **117**, A04102, doi:10.1029/2011JA017193

Millward, G.H., Rishbeth, H., Fuller-Rowell, T.J., Aylward, A.D., Quegan, S.,  Moffett, R.J.: 1996, Ionospheric F 2 layer seasonal and semiannual variations, *J. Geophys. Res*. **101**, 5149. doi:10.1029/95JA03343.

Ostrow, S. M., PoKempner, M.:1952, The differences in the relationship between ionospheric critical frequencies and sunspot number for different sunspot cycles, *J. Geophys. Res*. **57**, 473. doi:10.1029/JZ057i004p00473.

Phillips, M.L.: 1947, The ionosphere as a measure of solar activity, *Terr. Mag. Atmos. Electr*., 52, 321.  doi:10.1029/TE052i003p00321

Piggott, W.R., Rawer,K.: 1961, URSI Handbook of Ionogram Interpretation and Reduction (UAG-23A), Elsevier, Amsterdam/New York.

Rishbeth, H.: 1990, A greenhouse effect in the ionosphere? *Planet. Space Sci*. **38**, 945. doi:10.1016/0032-0633(90)90061-T

Roble, R.G., Dickinson, R.E.: 1989, How will changes in carbon dioxide and methane modify the mean structure of the mesosphere and thermosphere? *Geophys. Res. Lett*.**16**, 1441. doi: 10.1029/GL016i012p01441

Satterthwaite, F.E.: 1946, An approximate distribution of estimates of variance components, *Biometrics Bull*. **2**, 110. doi:10.2307/3002019.

Scott, C.J., Stamper, R.: 2015, Global variation in the long-term seasonal changes observed in ionospheric F region data, *Ann. Geophys*. **33**, 449.  doi:10.5194/angeo-33-449-2015

Scott, C.J., Stamper, R., Rishbeth, H.: 2014, Long-term changes in thermospheric composition inferred from a spectral analysis of ionospheric F-region data, *Ann. Geophys*. **32**, 113. doi:10.5194/angeo-32-113-2014

Sethi, N.K., Goel, M.K., Mahajan, K.K.: 2002, Solar Cycle variations of foF2 from IGY to 1990, *Annales Geophys.* **20**, 1677.

Shapley, A.H.: 1947, Reduction of sunspot-number observations, *Pub. Astron. Soc.Pacific* **61** (358) 13.

Smith, P., King, J.W. : 1981, Long-term relationships between sunspots, solar faculae and the ionosphere, *J. Atmos. Terr. Phys*. **43**, 1057.  doi: 10.1016/0021-9169(81)90020-9

Smith, E.J., Wolf, J.H.: 1976, Observations of interaction regions and co-rotating shocks between one and five AU: Pioneer 10 and 11, *Geophys. Res. Lett*. **3**, 137. doi: 10.1029/GL003i003p00137

Sobotka, M.: 2003, Solar activity II: Sunspots and pores, *Astron. Nachr.* **324**, 369. doi: 10.1002/asna.200310132

Solanki, S.K., Schüssler, M., Fligge, M.: 2000, Secular evolution of the Sun's magnetic field since the Maunder minimum, *Nature* **480**, 445. doi:10.1038/35044027.

Svalgaard, L.: 2011, How well do we know the sunspot number?, *Proc. Int. Astron. Union* **7**, 27. Cambridge University Press, Cambridge, UK. doi:10.1017/S1743921312004590.

Svalgaard, L., Schatten, K.H.: 2015, Reconstruction of the Sunspot Group Number: the Backbone Method, *Solar Phys.* (in press).

Trísková, L., Chum, J.: 1996,  Hysteresis in dependence of foF2 on solar indices, *Adv. Space Res*. **18** (6), 145.  doi:10.1016/0273-1177(95)00915-9

Ulich, T., Turunen, E.: 1997,  Evidence for long-term cooling of the upper atmosphere in ionosonde data, *Geophys. Res. Lett*, **24**, 1103.  doi: 10.1029/97GL50896

Usoskin, I.G., Kovaltsov, G.A., Lockwood, M., Mursula,  K., Owens, M.J., Solanki , S.K.: 2016,  A new calibrated sunspot group series since 1749: Statistics of active day fractions, *Solar Phys*., in press, doi: 10.1007/s11207-015-0838-1

Webb, D.F., Howard, R.A.: 1994, The solar cycle variation of coronal mass ejections and the solar wind mass flux, *J. Geophys. Res*. **99**, 4201. doi: 10.1029/93JA02742.

Welch, B.L.: 1947,  The generalization of "Student's" problem when several different population variances are involved, *Biometrika* **34**, 28.  doi:10.1093/biomet/34.1-2.28.

Willis, D.M., Wild, M.N., Warburton, J.S.: 2015, Re-examination of the Daily Number of Sunspot Groups for the Royal Observatory, Greenwich (1874 – 1885), *Solar Phys.* (in press)

Willis, D.M., Coffey, H.E., Henwood, R., Erwin, E.H., Hoyt, D.V., Wild, M.N., Denig, W.F.: 2013a, The Greenwich Photo-heliographic Results (1874 – 1976): Summary of the observations, applications, datasets, definitions and errors, *Solar Phys.* **288**, 117. doi: 10.1007/s11207-013-0311-y

Willis, D.M., Henwood, R., Wild, M.N., Coffey, H.E., Denig, W.F., Erwin, E.H., Hoyt D.V.: 2013b, The Greenwich Photo-heliographic Results (1874 – 1976): Procedures for Checking and Correcting the Sunspot Digital Datasets, *Solar Phys.* **288**, 141. doi: 10.1007/s11207-013-0312-x

Wolf, R.: 1861, Vortrag uber die Sonne und ihre Flecken, *Mitth. über die Sonnenflecken*, 12. doi: 10.3931/e-rara-3058

Zou, L., Rishbeth, H., Müller-Wodarg, I.C.F., Aylward, A.D., Millward, G.H., Fuller-Rowell, T.J., Idenden, D.W., Moffett, R.J.: 2000, Annual and semiannual variations in the ionospheric F2-layer.mI. Modelling, *Annales Geophys*. **18**, 927. doi:10.1007/s00585-000-0927-8, 2000